\documentclass[9pt,twocolumn,twoside]{pnas-new}
\usepackage{aas_macros}
\usepackage{amssymb}
\usepackage{breqn}
\usepackage{amsmath}
\usepackage{graphicx}

\templatetype{pnasresearcharticle}

\newcommand{\rstar}{$R_{\star}$ }
\newcommand{\mstar}{$M_{\star}$ }

\newcommand{\rprs}{$R_{p}/R_{\star}$ }
\newcommand{\sesinw}{$\sqrt{e} \sin \omega $ }
\newcommand{\secosw}{$\sqrt{e} \cos \omega $ }
\newcommand{\rhos}{$\rho_{\star}$ }
\newcommand{\ntargs}{163 }
\newcommand{\nsystems}{101 }
\newcommand{\nsingles}{67 }
\newcommand{\nmultis}{96 }
\newcommand{\nmultisys}{34 }
\newcommand{\ntwoplanet}{16 }
\newcommand{\nthreeplanet}{11 }
\newcommand{\nfourplanet}{3 }
\newcommand{\nfiveplanet}{4 }
\newcommand{\arcsec}{\mbox{$^{\prime\prime}$}}

\setboolean{displaywatermark}{false}

\title{The Orbital Eccentricity Distribution of Planets Orbiting M dwarfs}

\author[a,1]{Sheila Sagear}
\author[a]{Sarah Ballard} 

\affil[a]{Department of Astronomy, University of Florida,
211 Bryant Space Science Center,
P.O. Box 112055,
Gainesville, FL 32611}

\leadauthor{Sagear} 

\significancestatement{The orbital eccentricities of exoplanets orbiting M dwarf stars may significantly affect their habitability, but are unknown. We extract this eccentricity distribution using a sample of transiting planets orbiting M dwarfs detected by NASA’s Kepler Mission with stellar density measurements. We find planets in apparently single-transiting systems are typically more eccentric than planets in multiply-transiting systems. The single-transit data prefer an eccentricity model composed of two distinct dynamically warmer and cooler sub-populations over three single-component models. With known planet demographics, we estimate the eccentricity distribution for the population of early- to mid-M dwarf planets in the local neighborhood, with implications for planetary formation and follow-up observations. Comparing our findings with similar studies for Sunlike stars suggests common dynamical excitation mechanisms.}

\correspondingauthor{\textsuperscript{1}To whom correspondence should be addressed. E-mail: ssagear$@$ufl.edu}


\begin{abstract}
We investigate the underlying distribution of orbital eccentricities for planets around early-to-mid M dwarf host stars. We employ a sample of \ntargs planets around early- to mid-M dwarfs across \nsystems systems detected by NASA's \textit{Kepler} Mission. We constrain the orbital eccentricity for each planet by leveraging the \textit{Kepler} lightcurve together with a stellar density prior, constructed using metallicity from spectroscopy, $K_s$ magnitude from 2MASS, and stellar parallax from Gaia. Within a Bayesian hierarchical framework, we extract the underlying eccentricity distribution, assuming alternately Rayleigh, half-Gaussian, and Beta functions for both single- and multi-transit systems. We describe the eccentricity distribution for apparently single-transiting planetary systems with a Rayleigh distribution with $\sigma = {0.19}^{+0.04}_{-0.03}$, and for multi-transit systems with $\sigma = {0.03}^{+0.02}_{-0.01}$. The data suggest the possibility of distinct dynamically warmer and cooler sub-populations within the single-transit distribution: The single-transit data prefer a mixture model composed of two distinct Rayleigh distributions with $\sigma_1 = {0.02}^{+0.11}_{-0.00}$ and $\sigma_2 = {0.24} ^{+0.20}_{-0.03}$ over a single Rayleigh distribution, with 7:1 odds. We contextualize our findings within a planet formation framework, by comparing them to analogous results in the literature for planets orbiting FGK stars. By combining our derived eccentricity distribution with other M dwarf demographic constraints, we estimate the underlying eccentricity distribution for the population of early- to mid-M dwarf planets in the local neighborhood.
\end{abstract}

\dates{This manuscript was compiled on \today}
\doi{\url{www.pnas.org/cgi/doi/10.1073/pnas.XXXXXXXXXX}}

\begin{document}

\maketitle
\thispagestyle{firststyle}
\ifthenelse{\boolean{shortarticle}}{\ifthenelse{\boolean{singlecolumn}}{\abscontentformatted}{\abscontent}}{}

\dropcap{O}rbital eccentricity is a fundamental property of exoplanets. In addition to quantifying the current dynamical state, eccentricity also encodes information about planetary formation. Because eccentricity is associated with both variable insolation and tidal heating, it is also relevant to planetary habitability \cite{barnes_tidal_2009, jackson_tidal_2008, zanazzi_ability_2019}. For planets orbiting M dwarfs in particular, the relative proximity of the habitable zone to the star \cite{kopparapu_habitable_2013} means that even modest eccentricities can render planets inhospitable to life. For example, an Earth-like planet orbiting a 0.25 $M_{\oplus}$ star with an eccentricity $e>0.2$ could experience a ``tidal Venus" catastrophe, with tidal heating sufficient to evaporate a water ocean \cite{barnes_tidal_2013}. 

Investigating orbital eccentricity for planets orbiting M dwarfs is also pressing from a population standpoint: M dwarfs host planets smaller than Neptunes at a rate 3.5 times higher than Sunlike stars \cite{mulders_increase_2015}. They are themselves the most common type of star in our galaxy  (e.g. \cite{henry_solar_2004}; for review see \cite{mdwarfsaara}), and the planet-to-star size ratio renders them especially appealing for both detection and follow-up efforts \cite{tarter_reappraisal_2007}. All of these factors conspire to make them very likely targets for follow-up surveys to search for life \cite{tesstargets}. For these reasons, the eccentricities of planets around M dwarfs is of particular interest. The high eccentricity of the very first transiting planet discovered to orbit an M dwarf, GJ 436b \cite{maness_m_2007, deming_spitzer_2007}, is still mysterious fifteen years after its detection: while the Neptune-sized planet ought to have circularized over the age of its star, its high eccentricity persists without an as-yet detected perturber \cite{stevenson_lessigreaterhubble_2014}.

Much of our demographic information about planetary eccentricity, however, is obtained through radial velocity measurements and is focused upon Sunlike stars. Eccentricity is observationally correlated with planet multiplicity, with single-planet systems exhibiting higher eccentricities on average \cite{multiplicityrelation}. This relationship is moderated by both stellar metallicity and stellar binarity, however, with higher metallicity stars and stars in binary systems tending to host more eccentric planets \cite{mills_california-kepler_2019, dawson_giant_2013, moutou_eccentricity_2017}. Studies on the eccentricities of M dwarf planets tend to have smaller sample sizes, with patterns still newly emergent. Similarly to Sunlike stars \cite{moutou_eccentricity_2017}, stellar binarity appears to be a good predictor of increased eccentricity among M dwarf planets \cite{mann_gold_2017}. In addition, older M dwarfs host modestly eccentric planets as well as their younger counterparts \cite{veyette_chemo-kinematic_2018}.

A larger-scale study of the eccentricities of M dwarf planets would be helpful toward understanding their demographics. Yet, the gold standard for measuring orbital eccentricity, the radial velocity phase curve, is both time- and resource-intensive. It's also possible to measure eccentricities in closely-packed transiting planets in (near)resonant configurations that exhibit transit-timing variations \cite{hadden_densities_2014}, but only about a hundred are known. There exists another opportunity for transiting planets, even without radial velocity data or TTVs, to roughly constrain orbital eccentricity: the ``photoeccentric effect". First described by \cite{barnes_effects_2007} and \cite{ford_characterizing_2008} and applied by \cite{photoeccentric}, it relies on the relationship between the orbital speed (variable for an eccentric planet) and the transit duration. With strong constraints on stellar density, deviation from the standard circular orbital speed is encoded in a measurably short or long transit duration \cite{photoeccentric}. While first deployed for Hot Jupiters, the effect can be leveraged to constrain eccentricities for large samples of smaller planets, provided there are sufficiently informative density priors \cite{moorhead_2011, kane_evidence_2016, price_how_2015}. Previously, these samples have included stellar hosts with spectroscopically-constrained densities \cite{xie_exoplanet_2016, mills_california-kepler_2019, dong_warm_2021} or even asteroseismically constrained densities \cite{vaneylen2015, vaneylen}. While previous studies have mostly focused upon FGK dwarfs, \cite{mann_gold_2017} focused upon a sample of 8 M dwarfs. These stars possessed parallax measurements, enabling strong enough constraints on their densities to enable the eccentricity characterization of their planets. 

In this manuscript, we apply the photoeccentric method to the sample of M dwarf planetary hosts identified by NASA's \textit{Kepler} Mission, with the goal of estimating their underlying eccentricity distribution. In this work, we favor using Kepler data over TESS data because of its longer observation baseline and higher photometric precision. With the combination of spectroscopy and distance measurements from ESA's Gaia spacecraft, we demonstrate the ability to extract eccentricity constraints from the lightcurves of \ntargs planets. Within a hierarchical Bayesian framework, we go on to model the underlying eccentricity distribution for this sample. With demographic constraints for the mixture of M dwarf planetary systems in the Milky Way, we estimate the representative eccentricity distribution for the population of early- to mid-M dwarf planets, weighted by occurrence rate, in the local neighborhood.We note that references to ``single-transit systems" in this manuscript refer to systems with a single known transiting planet which may have other, non-transiting or undetectable planets.

\section*{Methods} \label{sec:methods}

To extract eccentricity measurements for the sample of \textit{Kepler} M dwarf planets, we take advantage of the so-called ``photoeccentric effect”, the phenomenon by which an eccentric planet’s transit duration (the time for a transiting planet to cross its stellar disk) differs from an analogous circular planet. The total transit duration $T_{14}$ and total transit duration from second to third contact $T_{23}$ are directly measurable from transit light curves, as well as the orbital period $P$, planet-to-star radius ratio $R_p/R_s$ (see \cite{Winn10_chapter} for additional description). \cite{photoeccentric} derived an expression involving $T_{14}$, $T_{23}$ and the density of the host star $\rho_{\star}$ as follows:

\begin{dmath}
    \rho_{\star} = g^{-3} \biggl ( \frac{2 \delta^{1/4}}{\sqrt{T_{14}^2 - T_{23}^2}} \biggr )^3  \frac{3 P}{G\pi^2},
    \label{eq:rhostar}
\end{dmath}

where $g$ is defined to be

\begin{equation}
    g(e,\omega) = \frac{1 + e \sin\omega}{\sqrt{1-e^2}}
    \label{eq:gdef}
\end{equation}

In this way, $\rho_{\star}$ is related to two quantities: a quantity dependent entirely on transit observables ($T_{14}$, $T_{23}$, $\delta$, and $P$), and a quantity $g$ that encodes eccentricity information. With prior information about  $\rho_{star}$, $g$ is in principle extractable. 

Using stellar metallicities and errors from the literature (compiled by \cite{mathur_revised_2017}), stellar parallaxes from Gaia \cite{Gaia1, Gaia2}, and $K_S$ magnitudes from 2MASS \cite{2MASS}, we calculate the stellar mass using the empirical, semi-model-independent $M_{K_S}$-$M_{\star}$-$[Fe/H]$ relation from \cite{mann_2019} and the stellar radius using the $M_{K_S}$, Fe/H, and \rstar relation from \cite{mann_2015}. We combine the mass and radius to constrain $\rho_{star}$. We use this independent constraint to construct a prior on $\rho_{star}$ and fit for transit observables, including the period, transit depth, impact parameter, eccentricity and longitude of periastron, for each planet in our sample. A detailed description of our methodology can be found in the SI Appendix.

\section*{Analysis and Results} \label{sec:analysis}

\begin{figure*}[!htbp]
    \centering
    \includegraphics[width=160mm, scale=0.1]{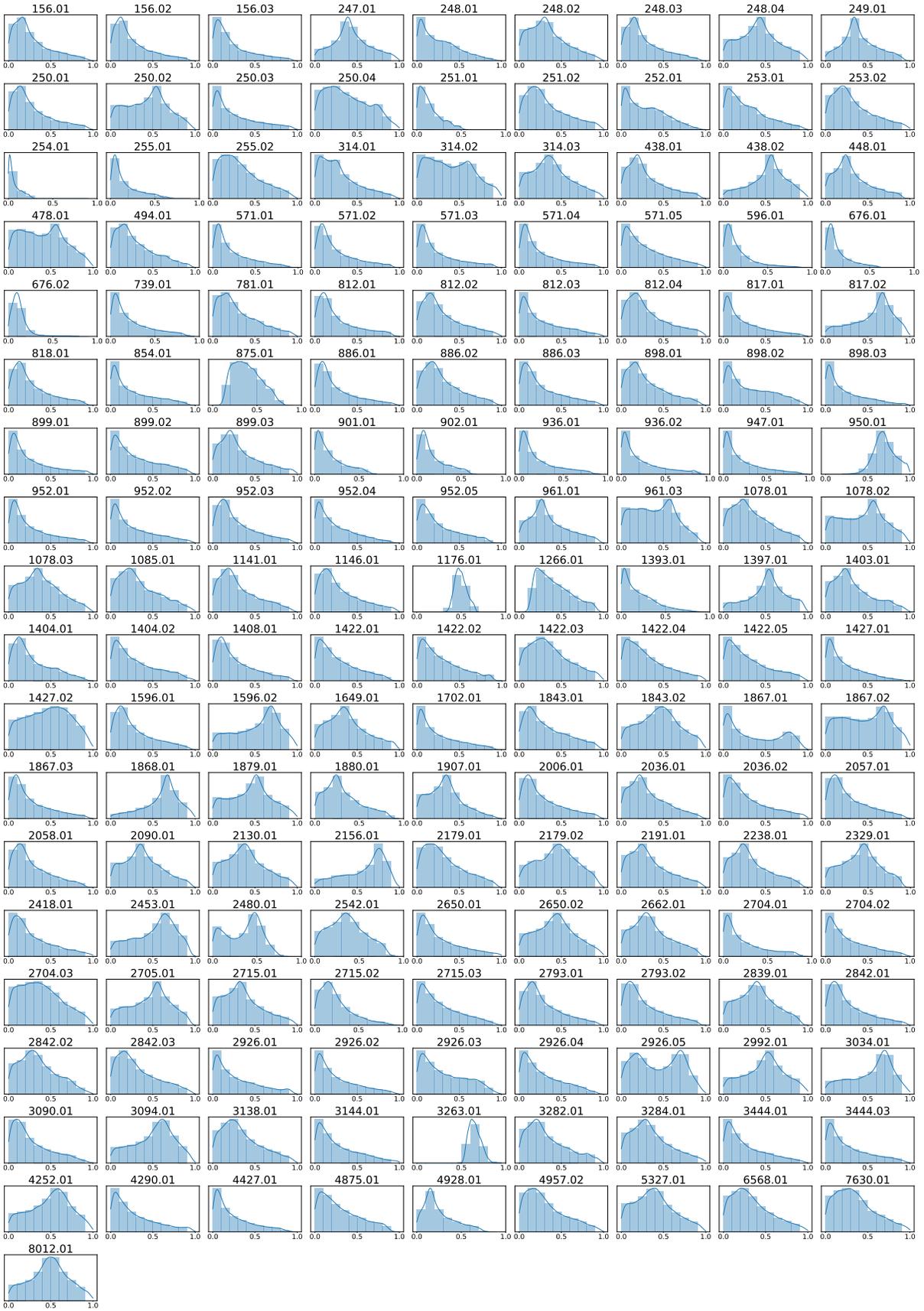}
    \caption{Individual $e$ posteriors for each planet, denoted by KOI. Bin widths are arbitrarily set to 0.1. The solid line represents a kernel density estimate of each distribution.}
    \label{fig:fam}
\end{figure*}

\begin{figure*}[!htbp]
  \includegraphics[width=1.0\linewidth]{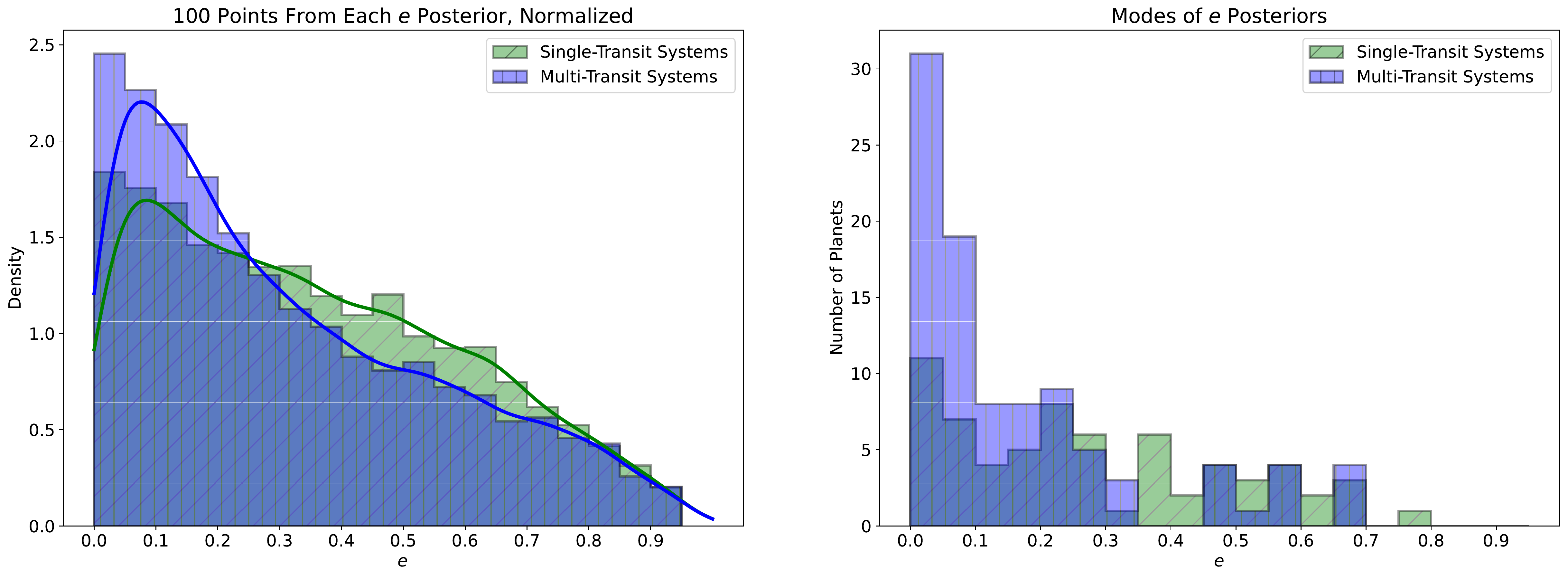}
  \caption{
  \textit{Left:} Normalized histogram of 100 points drawn from $e$ posteriors from each planet, separated into multi-planet systems (red) and single-planet systems (blue). \textit{Right:} Histogram of the modes of $e$ posteriors for each planet, separated into multi-planet systems (blue) and single-planet systems (green). Each bin width is set arbitrarily to $0.05$. Each bin width is set arbitrarily to $0.05$. A kernel density estimate is overplotted (blue and green lines for multi- and single-planet systems, respectively.)}
  \label{fig:histograms}
\end{figure*}

The individual $e$ posteriors for our sample are shown in Figure \ref{fig:fam}. Corner plots of the $R_p/R_{\star}$, $e$, $b$, and $\rho_{\star}$ posteriors for each fit are shown in the SI Appendix. The modes of each $e$ posterior and a combined, normalized histogram containing 100 points drawn from each $e$ posterior are shown in Figure \ref{fig:histograms}. The fit planet parameters are listed in Table \ref{tab:fitplanetparams}. Because the $e$ posterior is highly correlated with other free parameters, we report the mode of each $e$ posterior along with the $16^{th}$ and $84^{th}$ percentiles. With the photoeccentric effect, it is not always possible to precisely constrain $e$ for a single planet, especially for systems with few observed transits and long-cadence data. We note that even if the mode of a given $e$ posterior is greater than zero, it is often impossible to rule out $e=0$.

\begin{table*}
\resizebox{\textwidth}{!}{%
\begin{tabular}{cccccccccccc}
\hline
KOI & LC Days & SC Days & $R_p/R_{\star}$ & $b$ & $\rho_{\star}$ & $e sin \omega$ & $e cos \omega$ & $e$ & $\omega$ ($deg$) & $e_{pointest}$ \\
\hline
\vspace{0.01cm}
156.01 & 428 & 1228 & $0.022_{-0.001}^{+0.001}$ & $0.47_{-0.289 }^{+0.149}$ & $2.308_{-0.09 }^{+0.091}$ & $0.004_{-0.152}^{+0.12}$ & $0.0_{-0.31}^{+0.291}$ & $0.208_{-0.134}^{+0.303}$ & $4.294_{-151.351}^{+132.662}$ & 0.014 \\
156.02 & 428 & 1228 & $0.017_{-0.0}^{+0.001}$ & $0.402_{-0.25 }^{+0.174}$ & $2.308_{-0.093}^{+0.092}$ & $-0.001_{-0.169}^{+0.087}$ & $0.0_{-0.261}^{+0.283}$ & $0.184_{-0.12}^{+0.299}$ & $-1.408_{-144.783}^{+134.478}$ & 0.008 \\
156.03 & 428 & 1228 & $0.036_{-0.001}^{+0.001}$ & $0.629_{-0.074}^{+0.062}$ & $2.304_{-0.089}^{+0.088}$ & $-0.053_{-0.176}^{+0.018}$ & $0.001_{-0.252}^{+0.256}$ & $0.161_{-0.116}^{+0.317}$ & $-35.667_{-119.402}^{+84.544}$ & 0.007 \\
247.01 & 644 & 459 & $0.029_{-0.001}^{+0.001}$ & $0.43_{-0.287 }^{+0.257}$ & $4.138_{-0.157}^{+0.157}$ & $0.161_{-0.056}^{+0.322}$ & $0.008_{-0.447}^{+0.464}$ & $0.408_{-0.192}^{+0.257}$ & $62.626_{-81.534}^{+83.54}$ & 0.345 \\
248.01 & 396 & 987 & $0.039_{-0.001}^{+0.002}$ & $0.396_{-0.237}^{+0.273}$ & $3.497_{-0.146}^{+0.144}$ & $-0.095_{-0.33}^{+0.007}$ & $-0.002_{-0.259}^{+0.242}$ & $0.221_{-0.175}^{+0.282}$ & $-59.399_{-89.731}^{+82.097}$ & 0.002 \\
248.02 & 396 & 987 & $0.035_{-0.002}^{+0.002}$ & $0.593_{-0.352}^{+0.158}$ & $3.499_{-0.145}^{+0.144}$ & $0.016_{-0.209}^{+0.209}$ & $-0.001_{-0.364}^{+0.353}$ & $0.308_{-0.198}^{+0.271}$ & $11.274_{-148.538}^{+122.966}$ & 0.009 \\
248.03 & 396 & 987 & $0.026_{-0.001}^{+0.001}$ & $0.411_{-0.269}^{+0.17}$ & $3.503_{-0.141}^{+0.142}$ & $0.014_{-0.139}^{+0.119}$ & $0.0_{-0.306}^{+0.293}$ & $0.203_{-0.129}^{+0.301}$ & $11.936_{-157.136}^{+126.21}$ & 0.003 \\
248.04 & 396 & 987 & $0.024_{-0.001}^{+0.002}$ & $0.593_{-0.38 }^{+0.209}$ & $3.498_{-0.145}^{+0.147}$ & $0.095_{-0.172}^{+0.335}$ & $-0.001_{-0.437}^{+0.441}$ & $0.418_{-0.25}^{+0.248 }$ & $38.156 _{-150.858}^{+100.772}$ & 0.01 \\
249.01 & 799 & 607 & $0.04_{-0.001}^{+0.001}$ & $0.388_{-0.256}^{+0.249}$ & $7.835_{-0.333}^{+0.334}$ & $0.158_{-0.029}^{+0.291}$ & $0.005_{-0.418}^{+0.447}$ & $0.367_{-0.157}^{+0.258}$ & $63.688 _{-73.983}^{+82.384}$ & 0.318 \\
250.01 & 396 & 987 & $0.05_{-0.002}^{+0.002}$ & $0.511_{-0.279}^{+0.167}$ & $3.645_{-0.164}^{+0.165}$ & $-0.039_{-0.239}^{+0.077}$ & $0.003_{-0.278}^{+0.305}$ & $0.21 _{-0.138}^{+0.309 }$ & $-26.423_{-119.451}^{+134.593}$ & 0.006 \\
\vspace{0.01cm}
\end{tabular}}
\caption{Transit fit parameters. We report the median, $16^{th}$ and $84^{th}$ percentile for each parameter. Because the $e$ posteriors are rarely well-described by a Gaussian, we report the point estimate of $e$ as the value corresponding to the modal values of $e \sin \omega$ and $e \cos \omega$ \cite{zakamska_2011}. $e$, $\omega$, $e \sin \omega$ and $e \cos \omega$ are not sampled directly but calculated from $\sqrt{e} sin \omega$ and $\sqrt{e} cos \omega$. Only a portion of the table is shown here to demonstrate its form and function. The full table is available in machine-readable form as "Dataset S1" in the Data Supplements. The full posteriors are also available in the Data Supplements.}
\label{tab:fitplanetparams}
\end{table*}

With eccentricity posteriors in hand, we now investigate the likeliest parent distribution from which these eccentricities are drawn. This requires a careful accounting for detection bias. The higher likelihood of eccentric planets to transit was noted by \cite{barnes_effects_2007} and \cite{ford_characterizing_2008}, given the boost in transit probability possessed by the planet at periapse. After describing our method of accounting for bias, we then turn to the extraction of the underlying eccentricity distribution. In these sections, we consider several individual functions for modeling the distribution. We then consider the evidence for a mixture model of two distributions (we confirm our ability to accurately recover a known underlying eccentricity distribution of a synthetic sample of transiting planets in the SI Appendix). 

\subsection*{Inference of Parent Eccentricity Distribution} 
\label{sec:bayes}

 Following the method of \cite{vaneylen}, we calculate the likelihood of observing our set of $e$ posteriors, given a parent distribution model with parameters $\theta$:

\begin{equation}
    p(obs|\theta) = \frac{1}{N} \prod_{k=1}^{K} \sum_{n=1}^{N} \frac{p(e_{k}^{n}|\theta)}{p(e_{k}^{n}|\alpha)}, 
    \label{eq:epdf}
\end{equation}
where the number of samples from each posterior $N = 100$, the number of planets $K = $ \ntargs, and the prior with parameters $\alpha$ is defined by $p(e_{k}^{n}|\alpha)$ \cite{dfmlikelihood}. We assume that all planets in the population were detected, and the $e$ posteriors of all planets are independent. Eccentric planets are geometrically likelier to transit overall, due to the enhanced transit probability at to periapse ($\omega = 90^{\circ}$) \cite{barnes_effects_2007, beta1, beta2}. We use the notation $\hat{t}$ to indicate a planet that fully transits its star (having an impact parameter $b < 1$). We now take the non-uniform transit probability into account with the joint ($e$, $\omega$) prior given a transiting planet $\hat{t}$

\begin{equation}
    p(e, \omega | \hat{t}) = \frac{1+e \sin \omega}{1-e^2}.
\end{equation}

Our joint $\{e,w\}$ posteriors, without accounting for this effect, will be biased toward higher eccentricities than the true parent distribution. We account for this non-uniform transit probability by multiplying each eccentricity posterior distribution $p(e_{k}^{n}|\theta)$ by the reciprocal of the the prior $p(e, \omega | \hat{t})$. 

We evaluate Equation \ref{eq:epdf}, replacing the likelihood $p(e_{k}^{n}|\theta)$ with the appropriate likelihood function for each Rayleigh, half-Gaussian, and Beta distributions. We employ a uniform prior for all distribution parameters. 
According to dynamical simulation models, a Rayleigh distribution appropriately describes the eccentricity distribution of small planets \cite{zhou_2007, juric_2008}. We include the half-Gaussian and Beta distributions to enable a one-to-one comparison with \cite{vaneylen}. We include the complete likelihood functions in the SI Appendix.

We use a Markov Chain Monte Carlo (MCMC) analysis with the Python package \texttt{emcee} \cite{emcee}. The chains were run with 32 walkers for 10,000 steps each, and we discarded a burn-in phase of 1,000 steps. To demonstrate convergence, we calculate the autocorrelation time as a function of step number for each fit. The final autocorrelation time is smaller than the total number of steps divided by 50. We include the full analysis of autocorrelation times in the Data Supplements.

Here, we take the step of dividing the sample into singly- and multiply-transiting systems. Per \cite{vaneylen}, \cite{xie_exoplanet_2016}, and \cite{mills_california-kepler_2019} there is a strong possibility that the single transiting planet systems and multiple transiting planet systems will be drawn from distinct parent distributions. We are doubly motivated by the appearance by eye in Figure \ref{fig:histograms} of distinct distributions, when comparing the modes of the eccentricity posterior distributions for singles and multis.   

\subsubsection*{Single Function Model}

\begin{figure*}[!htbp]
\centering
\begin{tabular}{cc}
    \includegraphics[width=1.0\linewidth]{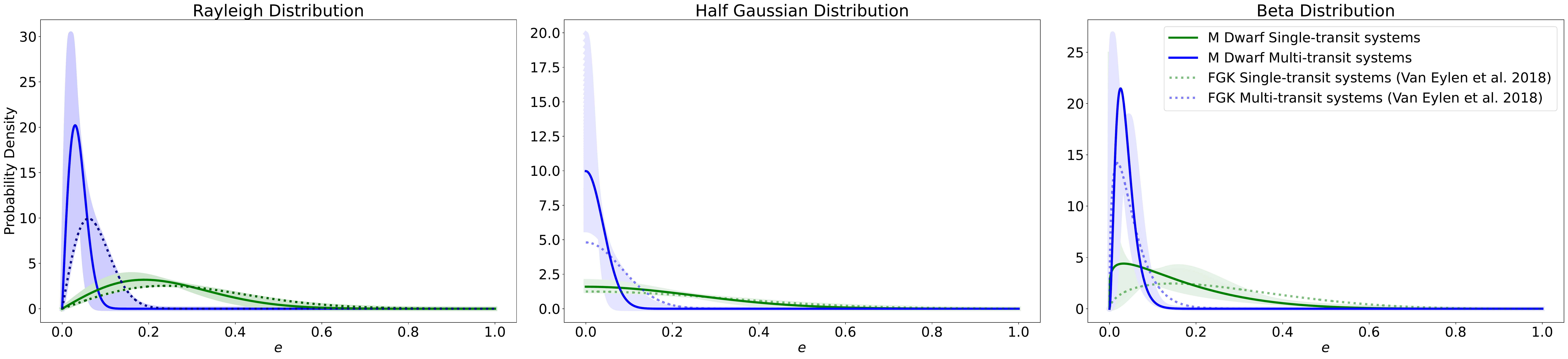} \\
\end{tabular}
\caption{\textit{Left:} Best-fit Rayleigh distributions for single-planet systems (green) and multi-planet systems (blue). \textit{Center:} Best-fit half-Gaussian distributions for single-planet systems (green) and multi-planet systems (blue). \textit{Right:} Best-fit Beta distributions for single-planet systems (green) and multi-planet systems (blue). The shaded regions represent the distributions corresponding to the $16^{th}$ and $84^{th}$ quantiles of the marginal eccentricity posteriors. The dotted lines represent best-fit distributions from a similar analysis by \cite{vaneylen} for FGK-type planet hosts.}
\label{fig:dists}
\end{figure*}

We first model the eccentricities with a single Rayleigh, single Beta, and then single half-Gaussian distribution, alternatively with the sample of singly-transiting planets, and then with the sample of multiply-transiting planets. For all trial functions, there is little overlap between the credible intervals for the best-fit model parameters for the populations of single and multiple transit systems. For example, the best-fit Rayleigh distribution for the single transiting planets is characterized by the modest eccentricity $\sigma = 0.19_{-0.03}^{+0.04}$, while the best-fit for the multiple transiting planets is much closer to circular, with $\sigma=0.03^{+0.02}_{-0.01}$. In Table \ref{tab:res}, we show the best-fit parameters for each distribution. Figure \ref{fig:dists} shows the best-fit Rayleigh, half-Gaussian, and Beta distributions and errors. 

We quantify the conclusion that the single-transiting and multi-transiting samples are best fit by two distinct distributions, rather than one distribution. We evaluate the probability ratio

\begin{equation}
    \label{eq:logr}
    R = \frac{P(D_1 D_2 | H_1)}{P(D_1 | H_0) P (D_2 | H_0)}
\end{equation}
\cite{hobson_2002, marshall_2006, feroz_2008} where $D_1$ and $D_2$ represent two subsets of data. The hypothesis $H_1$ states that the data is best described by a joint fit, and the hypothesis $H_0$ states that the data prefer separate models in different parts of parameter space. $P(D_1 D_2 | H_1)$ represents a joint fit to the entire dataset, while $P(D_1 | H_0)$ and $P(D_2 | H_0)$ represent individual fits to two data subsets. Where $P(D_1 | H_0)$ and $P(D_2 | H_0)$ are the likelihoods of the best-fit models for the singles and multis subsets, respectively, we find that $ln(R)$ is negative for the Rayleigh, Half-Gaussian, and Beta distribution models. For the Rayleigh fits, $log(R) = -7.09$; for the half-Gaussian fits, $log(R) = -6.72$; and for the Beta distribution fits, $log(R) = -7.78$. This shows strong evidence for $H_0$, in favor of two distinct models to describe the data \cite{feroz_2008}

\begin{table*}
\centering
\resizebox{\textwidth}{!}{%
\begin{tabular}{ccc}
\hline
Distribution & Parameters & Best-Fit Values \\
\hline
 Rayleigh & $(\sigma_{s}, \sigma_{m})$ & ${0.19}^{+0.04}_{-0.03}$, ${0.03}^{+0.02}_{-0.01}$ \\
 Half-Gaussian & $(\sigma_{s}, \sigma_{m})$ & ${0.25} ^{+0.06}_{-0.05}$, ${0.04}^{+0.03}_{-0.02}$ \\
 Beta & $(a_{s}, b_{s}), (a_{m}, b_{m})$ & (${1.18}^{+1.95}_{-0.61}$, ${6.34} ^{+5.94}_{-2.77}$); (${2.95}^{+2.34}_{-1.61}$, ${75.46}^{+17.68}_{-27.45}$) \\
 Mixture & $(\sigma_{1s}, \sigma_{2s}, f_{s}); (\sigma_{1m}, \sigma_{2m}, f_{m})$ & (${0.02}^{+0.11}_{-0.00}$, ${0.24} ^{+0.20}_{-0.03}$, ${0.86} ^{+0.03}_{-0.41}$); (${0.03}^{+0.02}_{-0.01}$, ${0.06} ^{+0.36}_{-0.02}$, ${0.99} ^{+0.01}_{-0.59}$) \\
 \vspace{0.01cm}
\end{tabular}}
\caption{Best-fit $e$ distribution parameters. For Rayleigh, half-Gaussian, and beta distribution models, the median$_{-16^{th}}^{+84^{th}}$ percentiles are shown. For the mixture model, the mode$_{-16^{th}}^{+84^{th}}$ percentiles are shown. For multi-transit systems, the best-fit $f$ is consistent with $1$, so the best-fit model is likely comprised entirely of a single Rayleigh distribution with $\sigma=\sigma_{1m}$. Therefore, $\sigma_{2m}$ is likely not meaningful.}
\label{tab:res}
\end{table*}

\subsubsection*{Mixture Model} \label{sec:mixture}

\begin{figure*}
  \vspace{0.3in}
  \hspace{0.3in}
  \centering
  \includegraphics[width=1.0\linewidth]{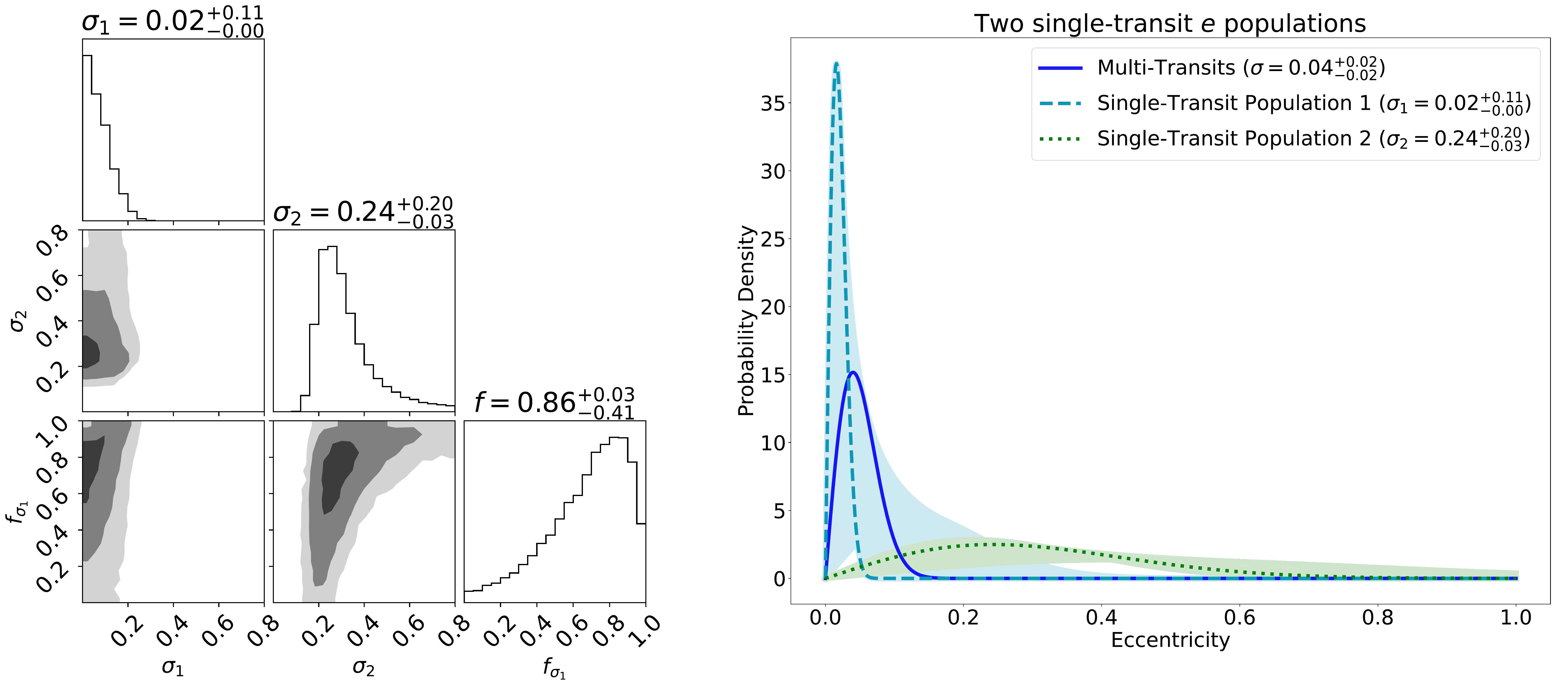}
  \caption{\textit{Left}: Corner plot of mixture model fit to single-transit $e$ distribution. $\sigma_1$ and $\sigma_2$ are the model parameters for two Rayleigh distributions that make up the mixture model. $f$ is the fractional contribution of $\sigma_1$ (if $f = 1$, the model is entirely defined by $\sigma_1$.) The statistical mode (${-16th, +84th}$) percentiles are shown as titles. The shaded regions represent the $1\sigma$, $2\sigma$ and $3\sigma$ posterior regions. The single-transit eccentricity distribution is best described by a mixed Rayleigh distribution with $\sigma_1 = 0.02$, $\sigma_2 = 0.24$, and $f = 0.86$. \textit{Right:} The best-fit parameters of the mixture model fit drawn as individual Rayleigh distributions: the light blue, dashed line represents the $\sigma_1$ single-transit model, and the green, dotted line represents the $\sigma_2$ single-transit model. We include the mixture model fit to the multi-transit eccentricity posteriors (blue solid line), which prefers a single Rayleigh distribution (with $f \approxeq 1$). The shaded regions represent the distributions corresponding to the $16^{th}$ and $84^{th}$ quantiles of the marginal eccentricity posteriors. The $\sigma_1$ single-transit model is consistent with the multi-transit model, and the $\sigma_2$ single-transit model peaks at higher eccentricity. This suggests that the single-transit sample is drawn from two distinct $e$ distributions: one of true multi-planet systems, and one of true single-planet systems with higher $e$.}
  \label{fig:Mixture}
\end{figure*}

The appearance of a distinct eccentricity parent distribution for singly- and multiply- transiting systems hints at a range of underlying dynamical temperatures among the exoplanetary systems \cite{tremaine_statistical_2015}. In this scenario, planetary systems presenting multiple transits possess lower mutual inclinations on average; by extension due to equipartition, these planets also possess lower eccentricities (e.g. \cite{fabrycky_architecture_2014}). \cite{ballard_kepler_2016} and \cite{dawson_correlations_2016} quantified the contribution of each of these hypothetical underlying populations to the \textit{Kepler} planetary yield. Some fraction of the time, \textit{bona fide} dynamically cool systems of multiple planets will present only a single transiting planet due to geometric probability. In this sense, the sample of single-transit systems can be understood to be "contaminated", in that it contains more than true dynamically warmer single-planet systems. Rather, it also includes a contribution from dynamically cooler multi-planet systems for which only one transit is observed. Indeed, dynamically cool but geometrically unlucky systems contribute to the population of singly-transiting systems at the level of $\sim$50\% \cite{ballard_kepler_2016}. We hypothesize that the single-planet eccentricity distribution can be modeled by one \textit{or more} Rayleigh distribution, and pose a test for this hypothesis as a model comparison between a one and two Rayleigh distribution model. Both model functions are described by the following expression, which assigns a weight $f$ to one of the Rayleigh distributions and correspondingly $1-f$ to the second:
\begin{equation}
    M(x) = f \times R_{\sigma_1}(x) + (1-f) \times R_{\sigma_2}(x)
    \label{eq:mixture}
\end{equation}
where $M(x)$ is the mixture model likelihood at a point $x$, and $R_{\sigma_n}(x)$ is the likelihood of a Rayleigh distribution with $\sigma_n$ at a point $x$. The mixture model parameters now correspond to two independently varying Rayleigh $\sigma$s ($\sigma_1$ and $\sigma_2$), and the relative weight of $\sigma_1$ to $\sigma_2$ ($f$). We observe that the one-Rayleigh model is simply a special case of Equation \ref{eq:mixture}; if $f=1$ or $f=0$,  $M(x)$ expresses a single Rayleigh distribution. We now evaluate the likelihood $p(obs|\theta)$, where $\theta$ now refers to $f$, $\sigma_{1}$, and $\sigma_{2}$. As before, we apply a uniform prior to all parameters. For the sake of removing redundancy, we require $\sigma_1 \le \sigma_2$. Table \ref{tab:res} shows the best-fit mixture model parameters for the single-transit eccentricity distribution. In Figure \ref{fig:Mixture}, we show the corner plot for the single-transit model fit and the best-fit mixture distributions. For single-transit systems, we find that the best two-Rayleigh model is characterized by $\sigma_1 = 0.02_{-0.00}^{+0.11}$, $\sigma_2 = 0.24_{-0.03}^{+0.20}$, and $f = 0.86_{-0.41}^{+0.03}$. The best single-Rayleigh model is the same as the one inferred in the preceding single-Rayleigh analysis: $\sigma = 0.19^{+0.04}_{-0.03}$ and $f=1.0$. 

We can now calculate the Bayesian Information Criterion (BIC):

\begin{equation}
    \textrm{BIC} = k \ln{n}-2\ln{\mathcal{L}},
\end{equation}

where $k$ corresponds to the number of parameters in our model, $n$ to the number of observations contained in $x_{n}$, and $\mathcal{L}$ to the peak likelihood of the model, $p(obs|\theta)$. A comparison between the BIC for the single-Rayleigh model and the BIC for the two-Rayleigh model will illuminate the preference of the data for one model over the over, and allow us to judge the merit of the two-Rayleigh model as a descriptor for our eccentricity distribution of singly-transiting planets.

We employ here a ready simplification for nested mixture models such as ours, in which one model is simply a special case of the generalized model. Both models were fit with the same parameters $\theta$, and therefore $k$ is the same; since we are employing the same data set to test both the one- and two-Rayleigh distributions, $n$ is also the same. We therefore find that the difference $D$ between the two BIC values is described by

\begin{equation}
    D = 2 \mbox{ln} \biggl( \frac{\text{Peak likelihood of two-Rayleigh model}}{\text{Peak likelihood of one-Rayleigh model}} \biggr)
\end{equation}
We find $D=7.17$, with the two-Rayleigh model corresponding to the lower BIC. When analyzing the eccentricity distribution for single-transit systems, the $\Delta BIC$ heuristic favors our model with two distinct distributions (where one closely matches the distribution of multi-transit eccentricities and one peaks at a higher $e$ of 0.24) over our model with one component. However, $\Delta BIC$ is biased to favor the more complex model, as it does not account for the value of the prior probabilities or the differing volume of the peaks in the posteriors for the two models. A proper Bayesian model comparison is beyond the scope of this paper.

We hypothesize the following physical interpretation: that the latter Rayleigh distribution may be considered as the underlying $e$ distribution for ``true" dynamically hotter single-planet systems. While this hypothesis has not been explicitly tested in this work, the dynamical heat of a system is in part determined by its eccentricity, and planetary systems with high eccentricities may be associated with higher mutual inclinations \cite{millholland2021}. This hypothesis will be explicitly investigated in a future work.

We apply the same methodology to the multi-transit systems and find a $D$ value of $-3 \times 10^{-6}$, indicating no strong preference between the one- and two-Rayleigh models.  We conclude that the eccentricities of multi-transit systems are drawn from a single distribution with no contamination, as expected. Similarly to previous analyses, we demonstrate the validity of our methodology by correctly recovering the parameters of a known mixture model in the SI Appendix.

\section*{Discussion} \label{sec:discussion}

We now consider our findings from the preceding section, within a context of other exoplanet eccentricity studies. While we cannot presently compare our findings to other larger-scale studies of the underlying M dwarf planet eccentricity distribution, we can compare to inferred eccentricity distributions of planets orbiting larger stars. \cite{vaneylen2015} and \cite{vaneylen} constrained the eccentricities of 66 multi-transit systems and 53 single-transit systems (respectively) around G-type stars, using a framework we have employed here. They found the best-fit half-Gaussian distribution for single- and multi-planet systems to be $\{\sigma_{single}, \sigma_{multi}\} = \{{0.32 ^{+0.06}_{-0.06}, 0.083^{+0.015}_{-0.02}\}}$. They found the best-fit Rayleigh distribution for single- and multi-planet systems to be $\{\sigma_{single}, \sigma_{multi}\} = \{{0.24 ^{+0.04}_{-0.04}, 0.061^{+0.01}_{-0.012}\}}$. Notably, \cite{vaneylen} found a significant difference between the underlying $e$ distributions for singly-transiting and multi-transiting systems. We note that ``singly-transiting" systems in \cite{vaneylen} refer to systems which have one known transiting planet, but may contain other non-transiting planets. There is significant overlap between the best-fit model parameters for our underlying eccentricity distributions and those of \cite{vaneylen}. The consistency of our results suggests that the underlying eccentricity distribution of small planets may dependent weakly on stellar spectral type, if at all. 

\subsection*{Consideration of Dynamical Mixture}

We find from our investigation that M dwarf exoplanets exhibit evidence for the two different underlying parent distributions in eccentricity: one dynamically cooler (associated with multiple transits), and one dynamically warmer (associated with single transits). This extends the similar findings for FGK dwarfs \cite{vaneylen, xie_exoplanet_2016, mills_california-kepler_2019} to later spectral types. There is also modest evidence to support the claim that the eccentricity distribution of singly-transiting planets is better described by contributions from both dynamically cold and warm contributions. We understand this finding in light of the geometric transit probability: some single transits are attributable to \textit{bona fide} dynamically warmer planets, while the remaining fraction are drawn from dynamically cool systems in which only one member transits. The degree to which the singly-transiting population is mixed (that is, the value of $f$) is of interest. While it peaks at a 90\% contribution from \textit{bona fide} dynamically cool systems, a 50/50 mixture is allowable at the $1\sigma$ level. This shows broad consistency with the $\sim$50\% predicted contribution of dynamically cool systems to the population of single transit hosts from \cite{ballard_kepler_2016} and \cite{dawson_correlations_2016}. It may alternately be the case that the two-population model of ``dynamically cooler" and ``dynamically warmer" planets we have employed here is overly simplified. \cite{he_architectures_2020} and \cite{zhu_about_2018} modeled the range of dynamical temperature in planetary systems as a single continuous distribution in $\{e,i\}$ space. Both demonstrated the consistency of that model with the observable properties of transiting planets as well. Indeed, for the sample of radial velocity planets with measured eccentricities, \cite{multiplicityrelation} showed a gradual increase with average eccentricity as the number of planets decreased. We have not tested whether the eccentricity measurements of our planetary sample are better modeled by a continuous distribution versus a bimodal one, but this may be possible with a larger sample. In any case, we find broad consistency between the relative contribution of dynamically warm and cool systems required to model our eccentricity measurements, as described by transit multiplicity distributions in other studies.

Whether the mixture of dynamical temperature is best modeled as a continuous or bimodal function, the mixture encodes information about the formation and subsequent evolution of planetary systems. It may be the case that the mixture is baked in during the earliest stages of formation. As \cite{he_architectures_2020} demonstrated, the continuous range in $\{e,i\}$ space that correctly replicates many observables of the \textit{Kepler} planet yield is predictable: planetary systems are clustered at the ``angular momentum deficit" stability limit as a natural outcome of pebble accretion.  The interaction between larger planetesimals with one another during the first 10 Myr of formation may also generate larger eccentricities and mutual inclinations, though this effect varies with the surface density distribution of planetesimals in the disk \cite{moriarty2016} and the presence of ambient gas \cite{dawson_correlations_2016}. \cite{pu__spacing_2015} demonstrated the plausibility of self-excitation of planetary systems, which are born only metastable at formation. The timescale for this excitation is 10s to 100s of Myr after formation \cite{pu__spacing_2015}. 

Planetary systems can also possess dynamically distinct component parts. \cite{zawadzki_migration_2022} showed that migration trapping can produce systems with dynamically cold inner planets, decoupled from dynamically warmer planets further from the star. Other studies have investigated the population of ultra-short-period (USP) exoplanets, which can be substantially misaligned with other planets orbiting the same star. In this case, it is the inner planets that often exhibit evidence for inclination excitation. \cite{dai_larger_2018} showed that innermost planets exhibit typical mutual inclinations of $7^{\circ}$ with planets further from the star. This is significant, as the further planets themselves show mutual inclinations of typically 2$^{\circ}$. The high mutual inclination of USP planets, compared with the other planets orbiting the same star, may also be due to tidal evolution \cite{millholland_formation_2020} or interaction with the quadrupole moment of the host star \cite{brefka_general_2021}. The relative fraction determined in this work of singly-transiting systems that are eccentric, versus those that are closer to circular, may be a useful distinguishing diagnostic between these scenarios. 

The fact that M dwarf systems tend to lack external giant planetary perturbers, as compared to FGK-type systems, is a suggestive hint \cite{mdwarfgiants}. \cite{vaneylen} considered the possibility that single-transit FGK dwarf systems may be more likely to be eccentric due to perturbations from outer giant planets. This effect was quantified in the simulations of \cite{becker_effects_2017}. The original flat configuration of a multi-planet system may be disturbed by a massive outer planet, exciting one or more from the transit geometry. Compact multi-planet systems may be more resistant to this effect \cite{laipu2017, pulai2018}. But because single- and multi-transit systems appear to have qualitatively similar eccentricity distributions for M and FGK dwarfs, our finding suggests the possibility of a dynamical mechanism that does not depend on the presence of giant perturbers. However, there exists significant uncertainty in the rate of giant planet occurrence around both FGK and M dwarfs, and the lack of stellar type dependence on eccentricity is not enough to confirm this suggestion. We warn that though the $e$ distributions for M and FGK dwarfs appear similar, the sample size in this work may not be large enough to prove statistically significant differences between the distributions. For \textit{giant} planets orbiting FGK dwarfs, there is strong evidence that the higher eccentricities associated with higher metallicity are driven by the presence of another giant planet \cite{dawson_giant_2013}. But the relationship between eccentricity and metallicity for small planets may require an interpretation that does not invoke the formation of (and subsequent agitation by) giant planets, given the similarity of the eccentricity distribution of small planets across spectral type.  \cite{mills_california-kepler_2019} found evidence that Kepler planets with high eccentricities preferentially occur around metal rich ([Fe/H] $>$ 0) stars. Indeed, with the same sample of \textit{Kepler} M dwarfs as we employ in this study, \cite{anderson_higher_2021} showed some evidence that multiple-transiting planet systems are metal poor compared to single-transiting planet systems, and even more so compared to field stars: this establishes a common relationship across spectral type that dynamically cool systems are likelier around metal-poor hosts. \cite{Lu_2020} provided a non-giant-planet interpretation of the metallicity trend with planet occurrence. They argued that the relationship between M dwarf metallicity and raw planet occurrence (higher for metal-rich stars) is evidence for \textit{planetesimal} accretion rather than pebble accretion. The extraction of the relationship between orbital eccentricity and stellar metallicity may be hindered for now by a relatively small lever arm in metallicity (95\% of the sample span a range of only -0.5 to 0.5 dex, per \cite{swift_characterizing_2015}). 

\cite{vaneylen} also consider the possibility of self-excitation as the cause of the difference in single- and multi-transiting eccentricity distributions, as formation conditions that cause high eccentricities also cause widely spaced orbits, larger mutual inclinations, and therefore low transit multiplicities for eccentric planets \cite{moriarty2016, dawson_correlations_2016}. In the case where self-excitation is significant, the resulting eccentricities would depend strongly on the solid surface density and the radial distribution of disk solids, which differ among stellar types \cite{moriarty2016, vaneylen}. In comparing to \cite{vaneylen} we find no significant dependence of the eccentricity distribution on stellar type, so we suggest that self-excitation may not be the most important process in exciting eccentricities. However, knowing the mutual inclination distribution of M dwarf planets is critical in quantitatively evaluating the importance of self-excitation. This concept will be investigated in a future work.

\subsection*{Full Underlying e distribution}

We now infer the eccentricity distribution for the general population of M dwarf systems, combining our findings with planet occurrence rates for single- and multi-planet systems in the literature. We caution that our findings extend to planets $\ge1.5R_{\oplus}$ with orbital periods $<200$ days, where \textit{Kepler}'s completeness is highest for M dwarfs \cite{Dressing15}, and that our results are most applicable for the early-to-mid spectral types targeted by \textit{Kepler}. \cite{muirhead_2015} found that $21^{+7}_{-5} \%$ of mid-M dwarf systems host compact multiple systems, a number consistent with the fraction among early M dwarfs from \cite{ballard_kepler_2016}. 

\begin{figure}[!htbp]
  \vspace{0.3in}
  \centering
  \includegraphics[width=1.0\linewidth]{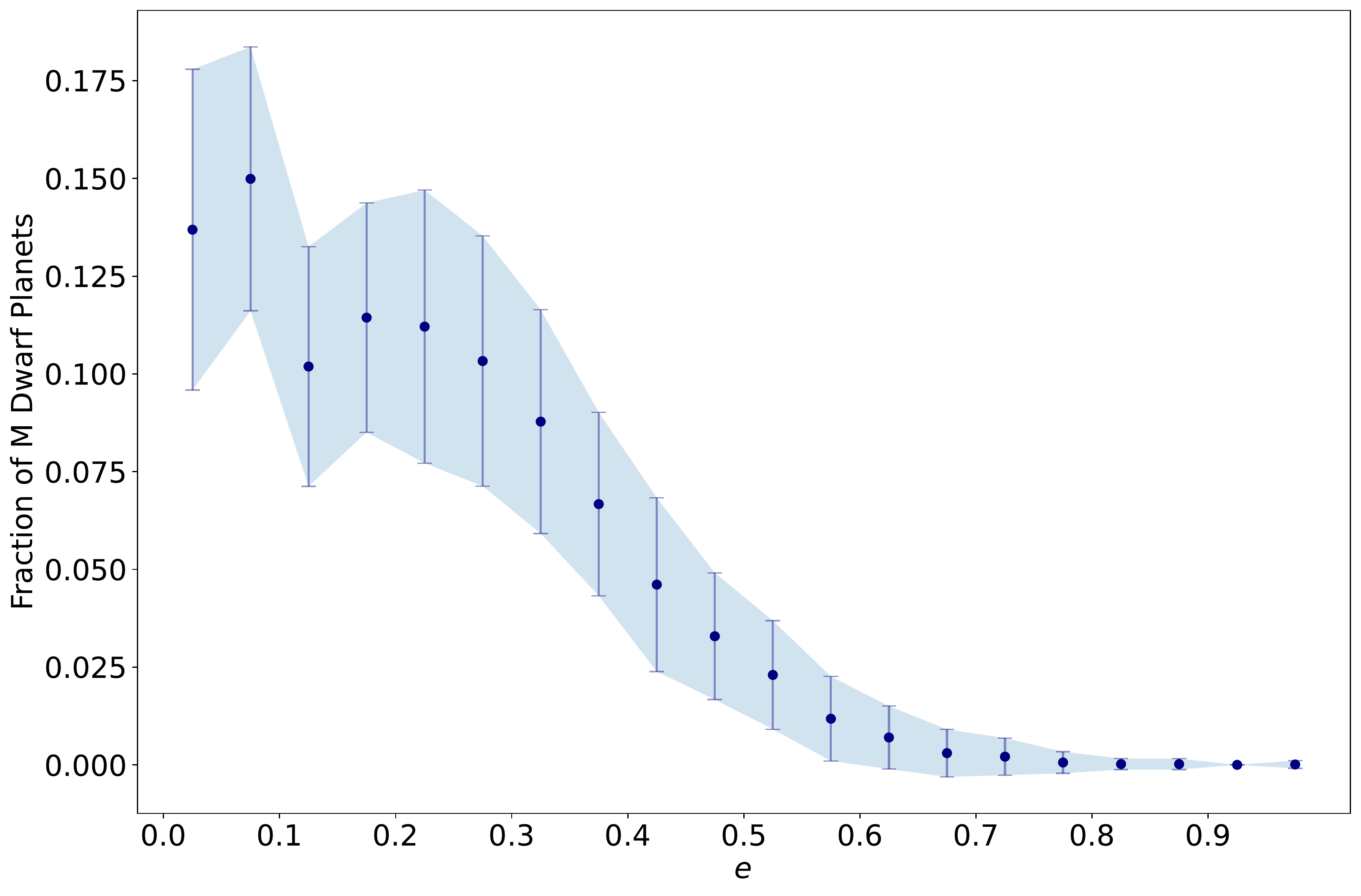}
  \caption{Fraction of all M dwarf planets with each eccentricity from simulations based on planet occurrence rates for single- and multi-planet systems from \cite{muirhead_2015}. We show the mean and 1$\sigma$ region from 100 simulations. Bins have widths of 0.05.}
  \label{fig:fulledist}
\end{figure}

To estimate the eccentricities of planets among a volume-complete sample of M dwarfs, we first assume that all M dwarfs host a planetary system of some kind, consistent with \cite{Dressing15}. The stellar sample in \cite{Dressing15} is skewed towards earlier-type M dwarfs relative to our sample and is potentially contaminated by late K dwarfs. Planet occurrence rate may therefore vary significantly across spectral types within M stars. However, occurrence rates broadly tend to increase with later spectral types \cite{howard_2012, mulders_increase_2015, mdwarfgiants}, so the assumption that each star in our sample hosts at least one planet appears to be valid. Among these, $21^{+7}_{-5} \%$ of stars in the sample are designated hosts to compact multiple systems \cite{muirhead_2015}, with the remainder hosting the dynamically warmer systems. For this analysis we assume that all compact multiple systems host exactly 5 planets, so we assume the number of single-planet hosts out of 100 M dwarfs is $100-21^{+7}_{-5}$. This number must be higher to explain systems like TRAPPIST-1 \cite{Gillon17}, and indeed it lies at the low end of posterior for number of planets per dynamically-cool-host, so in this sense our distribution is a lower-limit at low eccentricities. For 100 iterations, we draw the fraction of M dwarfs compact multi hosts from an asymmetric Gaussian with $\mu=0.21$, $\sigma_u = 0.07$ and $\sigma_l = 0.05$. We calculate the number of compact multiple and single-planet hosts out of 1000 planets for each draw. We take the eccentricity distribution for compact multiple systems to be a Rayleigh distribution with $\sigma_m = 0.03$, and for single-planet systems a Rayleigh distribution with $\sigma_s = 0.24$, according to the mixture model fits in the Results \& Analysis section. We draw eccentricities from the single- and multi-planet distributions corresponding to the fraction of single- and multi-planet systems in each iteration. The resulting group of 10,000 eccentricities reflects the complete underlying eccentricity distribution for M dwarf planets in a volume-complete sample (Figure \ref{fig:fulledist}). In this calculation, we have made simplifying assumptions: we have employed a single fiducial template for the number of planets in dynamically cool and warm systems, and have not folded the error in our measurements of $\sigma_m$ and $\sigma_s$ into our analysis; these quantities contribute much less to the uncertainty on the resulting $e$ distribution, however, than the uncertainty of the compact multi rate. 

From this distribution, it is clear that low orbital eccentricities ($e<0.1$) are not the norm among a typical sample of planets with periods $P<200$ days. While compact multiple planetary systems are the minority of M dwarf planetary systems as a whole, the larger number of planets-per-star among those systems does boost the occurrence of planets with nearly-circular orbits. Between 21-36\% of planets, with the 1$\sigma$ confidence interval, possess eccentricities $<0.1$ within the period and radius ranges where $\textit{Kepler}$ is most complete.

\section*{Conclusions} \label{sec:conclusion}

We constrained the eccentricities of \ntargs planets orbiting \nsystems M dwarfs. We performed our measurements via the `photoeccentric method', combining stellar densities for the \nsystems stars (derived from a combination of spectroscopy, \textit{Gaia} parallaxes, and 2MASS magnitudes), with transit durations from \textit{Kepler} light curves.  We employ the resulting $e$ posteriors within a hierarchical Bayesian framework to infer the underlying $e$ distribution for planets orbiting early-to-mid M-dwarfs, considering a variety of functional forms including a mixture models. We summarize our findings as follows:
\begin{itemize}
\item The eccentricities of single-transit and multi-transit systems are likely drawn from distinct underlying parent distributions. The eccentricity distribution for single-transit systems is best described by models that peak at higher $e$ than for multi-transit systems.

\item We find modest evidence that the single-transit population is best described with a dynamical mixture model, with dynamically warmer and dynamically cooler populations. We conclude that the sample as a whole is best modeled as a mixture of Rayleigh distributions: one peaking at $\sigma_2 = {0.21} ^{+0.28}_{-0.01}$, and the other at $\sigma = {0.04}^{+0.02}_{-0.02}$. The data for the single-transiting systems favor the dynamical mixture model over the single-population Rayleigh model with 7:1 odds.
\item The inferred parent distributions in orbital eccentricity for single- and multi-transit M dwarf systems are similar to analogous distributions for FGK dwarfs from the literature. Because M dwarfs tend to lack external giant planets when compared to larger stars, our findings favor a interpretation for dynamical excitation that does not require the presence of giant perturbers. In this sense, the eccentricity-metallicity relation for small planets (by which metal-poor stars tend to host lower eccentricity planets) may reflect a relationship other than metallicity's impact upon pebble accretion or planetesimal accretion early on, or self-excitation by neighboring small planets later in the system's lifetime.
\item We present an estimate of the underlying intrinsic $e$ distribution for the population of early- to mid-M dwarf planets in the local neighborhood with radii $>1.5R_{\oplus}$ and with periods $<200$ days, by combining our findings with other M dwarf planetary demographic constraints. Assuming the \textit{Kepler} sample is representative of typical early-to-M dwarfs in the galaxy, this distribution may typify eccentricities for planets orbiting small stars in the Milky Way.
\end{itemize}
The underlying eccentricity distributions presented here may be applicable for transit fit priors for small transiting planets in future studies. While our per-planet eccentricity constraint is quite coarse, these individual eccentricity posteriors may be useful toward target selection for follow-up observations. Furthermore, because our per-planet eccentricities are not well constrained, it is challenging to comment on the effects on habitability for individual planets. However, we contend that “compact multiple” systems may be the best place to search for habitable planets, as they appear more likely to have near-circular orbits.

Expanding this analysis using data from other surveys is promising. Applying our methods to measure the eccentricities of M dwarf planets observed by the Transiting Exoplanet Survey Satellite (TESS) \cite{ricker_2014} is also feasible, though this may prove to be more challenging with TESS due to the shorter observation baseline. Data from the PLAnetary Transits and Oscillations of stars (PLATO) Mission \cite{rauer_2014}, expected to launch in 2026, could prove to be useful for expanding this work due to its planned high precision and long observation baseline.

\subsection*{Data Availability}
All codes and data used in this manuscript are publicly available on Zenodo with the DOI 10.5281/zenodo.7731019 and on GitHub at https://github.com/ssagear/photoeccentric.

\showmatmethods{}

\acknow{We are grateful to Andrew Mann for helpful guidance on the transit fitting process. We thank Christopher Lam for thoughtful feedback on this manuscript. We thank the anonymous referees for carefully reviewing this manuscript and offering suggestions that improve the quality of this work.
This paper includes data collected by the Kepler mission and obtained from the MAST data archive at the Space Telescope Science Institute (STScI). Funding for the Kepler mission is provided by the NASA Science Mission Directorate. STScI is operated by the Association of Universities for Research in Astronomy, Inc., under NASA contract NAS 5–26555.
This work has made use of data from the European Space Agency (ESA) mission
{\it Gaia} (\url{https://www.cosmos.esa.int/gaia}), processed by the {\it Gaia}
Data Processing and Analysis Consortium (DPAC,
\url{https://www.cosmos.esa.int/web/gaia/dpac/consortium}). Funding for the DPAC
has been provided by national institutions, in particular the institutions
participating in the {\it Gaia} Multilateral Agreement.
This publication makes use of data products from the Two Micron All Sky Survey, which is a joint project of the University of Massachusetts and the Infrared Processing and Analysis Center/California Institute of Technology, funded by the National Aeronautics and Space Administration and the National Science Foundation.
This research has made use of the NASA/IPAC Infrared Science Archive, which is funded by the National Aeronautics and Space Administration and operated by the California Institute of Technology.
This research has made use of the Exoplanet Follow-up Observation Program (ExoFOP; DOI: 10.26134/ExoFOP5) website, which is operated by the California Institute of Technology, under contract with the National Aeronautics and Space Administration under the Exoplanet Exploration Program.
This work made use of the gaia-kepler.fun crossmatch database created by Megan Bedell.
This work made use of the following facilities: Kepler, Gaia, 2MASS, IRSA, Exoplanet Archive, ExoFOP.}

\showacknow{}

\newpage
\onecolumn

\begin{center}
\includegraphics[width=9.95cm]{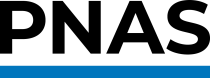}
\end{center}

\vskip45pt
\begingroup
\raggedright
{\Huge\sffamily\bfseries Supporting Information for\par}
\bigskip
{\LARGE\sffamily\bfseries The Orbital Eccentricity Distribution of Planets Orbiting M dwarfs \par}
\bigskip
{\sffamily\bfseries Sheila Sagear and Sarah Ballard \par\bigskip Sheila Sagear. \par Email: ssagear@ufl.edu \par}
\endgroup
\bigskip
\section*{This PDF file includes:}
\begin{list}{}{%
\setlength\leftmargin{2em}%
\setlength\itemsep{0pt}%
\setlength\parsep{0pt}}
\item Supporting text
\item Figs.~S1 to S4
\item Tables S1 to S3
\item SI References
\end{list}

\newpage
\onecolumn

\section*{Methods} \label{sec:SI_methods}

In this section, we detail our methodology for extracting eccentricity measurements from the sample of known \textit{Kepler} M dwarf planets. We first describe this sample, the lightcurve preparation process, and how we obtain our stellar density priors from this sample. We summarize the ``photoeccentric" formalism and our fitting pipeline (note that we quantify our sensitivity as a function of $\{e,w\}$ parameter space, by injecting and recovering these parameters from synthetic lightcurves later in the SI Appendix).

\subsection*{Stellar and Planetary Sample}
\label{sec:SI_sample}

We employ the stellar properties compiled in Kepler DR25 \cite{mathur_revised_2017}, which are listed in NASA's Exoplanet Archive Cumulative KOI table \cite{KOICumulative}. We take the effective temperature ($T_{\mbox{eff}}$) and stellar metallicity ([Fe/H]) from Kepler DR25. The effective temperatures and metallicities were compiled from the literature by \cite{mathur_revised_2017} (see Table 3 of \cite{mathur_revised_2017} for a list of provenances for each star). These stellar properties are not calculated by interpolating stellar isochrones.

To select our sample, we first remove all KICs with stellar host effective temperatures greater than 4000 K and listed disposition scores below 0.5 \cite{thompson_2018}. From this subset, we exclude KOIs with an Exoplanet Archive Disposition of "False Positive", according to the tests performed by \cite{batalha_2013}.

We cross-match these KICs with Gaia IDs using the $1 \arcsec$ radius \texttt{gaia-kepler.fun} crossmatch database. For KICs without $1 \arcsec$ database entries, we use the cross-match from the analogous $4$\arcsec radius database if there is only one entry. For confirmed planets with no entries in either database, or with no entries in the $1 \arcsec$ database but multiple entries in the $4 \arcsec$  database, we take the Gaia ID listed in the Exoplanet Archive's Planetary Systems table \cite{ConfirmedPlanets}. KOIs 1681, 2626, 3010, and 6276 were further excluded because they have Gaia IDs from one of these sources, but no parallax information. KOI 2862 was excluded because we found no matching Gaia ID from any of these sources. The parallax for KOI 1422 was taken from Gaia DR3, while all other parallaxes were taken from Gaia DR2.

We then cross-match these KICs with the 2MASS All-Sky Point Source Catalog using a $2 \arcsec$  cone search on the NASA/IPAC Infrared Science Archive \cite{2MASS}. KOIs 1201 and 1725 were further excluded because the 2MASS data for these targets were incomplete. We exclude KOIs 7408 and 8007 because their impact parameters listed in the Exoplanet Archive were greater than 1.2 \cite{KOICumulative}. We remove KOIs 605, 3497 and 7791 because our stellar mass calculations do not yield valid results for these stars, as we describe in the Calculation of Stellar Densities section. Finally, we exclude five KOIs (961.02, 4419.01, 6863.01, 7793.01, and 8037.01) because the computational needs of their transit fits significantly exceed the needs of the rest of the sample. We do not expect that excluding these KOIs significantly affects our results, given the size of our sample.

The final sample includes a total of \ntargs KOIs: \nsingles single-transit systems and \nmultis planets or candidates across \nmultisys multi-transit systems. We include \ntwoplanet two-planet systems, \nthreeplanet three-planet systems, \nfourplanet four-planet systems, and \nfiveplanet five-planet systems. We note that for one three-planet system (the KOI 961 system), we only include two out of three confirmed planets. The sample includes both confirmed planets and planet candidates. There are a total of 25 planet candidates in the sample with 16 candidates in single-transit systems, 8 candidates in two-planet systems, and one candidate in a five-planet system. Given the proportion of planet candidates to confirmed planets and our sample vetting process, we expect that this is a high-fidelity sample which is unlikely to contain significant numbers of false positives.

\subsection*{Lightcurve Preparation} 
\label{sec:SI_lightcurveprep}
We downloaded all available long- and short-cadence Kepler light curves for our sample from the Mikulski Archive for Space Telescopes (MAST). Where both long- and short-cadence data are available for a single quarter, we take the short-cadence data. For each target, we normalize the light curve data to 1 and stitch quarters together, preserving the original time stamps. We estimate the transit midpoints at the times observed using the transit start time and orbital period obtained from \cite{KOICumulative}.

To prepare the lightcurves for modeling, we first remove out-of-transit data. We find the closest flux point to each injected transit midpoint and isolate a window of 4 hours plus one half of the transit duration published in \cite{KOICumulative} before and after the transit midpoint. For KOIs 902.01, 2418.01, 2992.01, and 3263.01, we increased the length of the added baseline window to 8 hours before and after the transit (instead of 4 hours) due to their comparatively long transit durations. We discard the out-of-transit data. We fit a cubic model to the outer 2.5 hours of each transit segment. We subtract the cubic model from the entire transit segment. 


\subsection*{Calculation of Stellar Densities} 
\label{sec:SI_stellardensity}

We summarize here our calculation of the stellar density, which will ultimately apply as a prior during the lightcurve fit. The fundamental properties of M dwarfs are difficult to accurately extract from spectroscopy alone \cite{mann_2013a}, so we use the empirically driven method of \cite{mann_gold_2017}. For each star, we take stellar metallicities and errors from the literature (compiled by \cite{mathur_revised_2017}), stellar parallaxes from Gaia \cite{Gaia1, Gaia2}, and $K_S$ magnitudes from 2MASS \cite{2MASS}. We first calculate the absolute magnitude $M_{K_S}$ using the 2MASS $K_S$ magnitudes and parallaxes from Gaia. We then calculate the radius \rstar for each sample star with the $M_{K_S}$, Fe/H, and \rstar relation from \cite{mann_2015}. \cite{mann_2015} investigated the significance of correlations between $M_\star$ and $R_\star$, and demonstrated that these relations properly reproduce the covariance between mass and radius.

Next, we calculate stellar masses using the empirical, semi-model-independent \boldmath{$M_{K_S}$-$M_{\star}$-$[Fe/H]$} relation from \cite{mann_2019} using the \texttt{M_-M_K-} python software. The empirical relation was derived according to a sample of nearby M dwarfs with $4 < M_{K_s} < 11$ and metallicities of $-0.6 < [Fe/H] < 0.4$. We have excluded any stars from the sample which fall outside this range of $M_{K_s}$. There are six stars in our sample which have $4 < M_{K_s} < 11$ and $[Fe/H] =< 0.4$, but we include these stars in our analysis because the role of metallicity in the empirical relation is much less significant than the $M_{K_s}$ magnitude \cite{mann_2019}. The  empirical relation is best calibrated in the region between $-0.4 < [Fe/H] < 0.3$, and $90\%$ of our sample falls within this region. The empirical relation is best calibrated between $4.5 < M_{K_s} < 10.5$, and $88\%$ of our sample have $M_{K_s}$ greater than 4.5. Because $M_{K_s}$ has the strongest influence on the mass calculation, we urge caution in interpreting the calculated masses and resulting eccentricities of systems with $M_{K_s} < 4.5$ (Table \ref{tab:SI_calcstellarparams}).

We combine \rstar and \mstar to calculate the stellar density for each star. The literature stellar parameters we used in these calculations are listed in Table \ref{tab:SI_litstellarparams}. The calculated stellar parameters are listed in Table \ref{tab:SI_calcstellarparams}. The \cite{mann_2019} $M_{K_S}$, Fe/H, and \rstar relation should be restricted to main-sequence stars with $4 < M_K < 11$. At this stage, we exclude KOI systems 605, 3497 and 7791 because $M_K < 4$ for these stars. We note that where our sample includes KOIs around binary systems, we assume that the planet transits the primary star.

\begin{table} 
\centering
\resizebox{\textwidth}{!}{%
\begin{tabular}{cccccccccc}
\hline
KIC & KOI & Kepler Name & Gaia ID & 2MASS ID & T$_{eff}$ $(\mathrm{K})$ & [Fe/H] & $\pi (\mathrm{{}^{\prime\prime}})$ & $K_s$ \\
\hline
10925104 & 156 & Kepler-114 & 2128939873302216320 & 19362914+4820582 & 3980$\pm$79 & -0.2$\pm$0.15 & 0.003824$\pm$1.9e-05 & 11.37$\pm$0.02 \\
11852982 & 247 & Kepler-1712 & 2132326747070439296 & 18595966+5008484 & 3732$\pm$79 & 0.02$\pm$0.15 & 0.006371$\pm$1.9e-05 & 11.12$\pm$0.02 \\
5364071 & 248 & Kepler-49 & 2053523271244105216 & 19291070+4035304 & 3834$\pm$81 & -0.02$\pm$0.15 & 0.003185$\pm$2.6e-05 & 12.38$\pm$0.02 \\
9390653 & 249 & Kepler-504 & 2107186134525176960 & 18594123+4558206 & 3547$\pm$75 & -0.14$\pm$0.15 & 0.010065$\pm$3.1e-05 & 11.16$\pm$0.03 \\
9757613 & 250 & Kepler-26 & 2107317358665730688 & 18594583+4633595 & 3879$\pm$81 & -0.12$\pm$0.15 & 0.002954$\pm$2.7e-05 & 12.63$\pm$0.03 \\
10489206 & 251 & Kepler-125 & 2086439488284337536 & 19530194+4736178 & 3808$\pm$80 & -0.06$\pm$0.15 & 0.005425$\pm$2.9e-05 & 11.68$\pm$0.02 \\
11187837 & 252 & Kepler-1663 & 2129578826996503040 & 19213636+4849213 & 3744$\pm$79 & 0.06$\pm$0.15 & 0.002931$\pm$3e-05 & 12.55$\pm$0.03 \\
11752906 & 253 & & 2132144056342471168 & 19021784+4957441 & 3757$\pm$78 & 0.48$\pm$0.1 & 0.003091$\pm$3.5e-05 & 12.29$\pm$0.04 \\
5794240 & 254 & Kepler-45 & 2053562475706063744 & 19312949+4103513 & 3793$\pm$80 & 0.32$\pm$0.15 & 0.00259$\pm$4.3e-05 & 12.89$\pm$0.03 \\
7021681 & 255 & Kepler-505 & 2102511874378223360 & 19112594+4232334 & 3780$\pm$80 & -0.02$\pm$0.15 & 0.0035$\pm$2.2e-05 & 12.08$\pm$0.02 \\
\vspace{0.01cm}
\end{tabular}}
\caption{Literature stellar parameters. Gaia IDs are crossmatched with KICs using the \texttt{gaia-kepler.fun} crossmatch database. Effective temperatures and stellar metallicities are taken from \cite{mathur_revised_2017}. Parallaxes are taken from Gaia crossmatched data. $K_s$ magnitudes are taken from 2MASS crossmatched data. Only a portion of the table is shown here to demonstrate its form and function. The full table is available in machine-readable form as "Dataset S2" in the Data Supplements.}
\label{tab:SI_litstellarparams}
\end{table}

\begin{table}
\centering
\begin{tabular}{ccccc}
\hline
KOI & $M_{\star}$ ($M_{\odot}$) & $R_{\star}$ ($R_{\odot}$) & $M_{K_s}$ & $\rho_{\star}$ ($\rho_{\odot}$) \\
\hline
156 & 0.7065 $\pm$ 0.0195 & 0.7557 $\pm$ 0.0072 & 4.2787 $\pm$ 0.0246 & 1.6377 $\pm$ 0.0652 \\
247 & 0.5803 $\pm$ 0.0146 & 0.5827 $\pm$ 0.0057 & 5.1449 $\pm$ 0.0227 & 2.9341 $\pm$ 0.1123 \\
248 & 0.6166 $\pm$ 0.016 & 0.6285 $\pm$ 0.0068 & 4.8982 $\pm$ 0.029 & 2.4855 $\pm$ 0.1035 \\
249 & 0.4164 $\pm$ 0.0108 & 0.4218 $\pm$ 0.0049 & 6.1691 $\pm$ 0.028 & 5.5554 $\pm$ 0.2411 \\
250 & 0.6046 $\pm$ 0.0156 & 0.6158 $\pm$ 0.0073 & 4.9819 $\pm$ 0.0327 & 2.591 $\pm$ 0.1141 \\
251 & 0.5484 $\pm$ 0.0137 & 0.5484 $\pm$ 0.0052 & 5.354 $\pm$ 0.0222 & 3.3274 $\pm$ 0.126 \\
252 & 0.6182 $\pm$ 0.0161 & 0.6285 $\pm$ 0.0082 & 4.8869 $\pm$ 0.038 & 2.4932 $\pm$ 0.1179 \\
253 & 0.638 $\pm$ 0.0187 & 0.6437 $\pm$ 0.0089 & 4.7404 $\pm$ 0.0449 & 2.3946 $\pm$ 0.1209 \\
254 & 0.6068 $\pm$ 0.0173 & 0.6079 $\pm$ 0.0095 & 4.9588 $\pm$ 0.0467 & 2.7051 $\pm$ 0.1475 \\
255 & 0.6306 $\pm$ 0.0159 & 0.6469 $\pm$ 0.0061 & 4.8002 $\pm$ 0.0227 & 2.3304 $\pm$ 0.0879 \\
\hline
\end{tabular}
\vspace{0.01cm}
\caption{Calculated stellar parameters. Stellar masses are calculated using the $M_{K_S}$-$M_{\star}$ relation from \cite{benedict_solar_2016} with 2MASS $K_S$ magnitudes and Gaia parallaxes. Stellar radii are calculated using the $M_{K_S}$, Fe/H, and $R_{\star}$ relation from \cite{mann_2015} with 2MASS $K_S$ magnitudes and Gaia parallaxes. Only a portion of this table is shown here to demonstrate its form and content. The full table is available in machine-readable form as "Dataset S3" in the Data Supplements.}
\label{tab:SI_calcstellarparams}
\end{table}

\subsection*{Photoeccentric Effect Pipeline}
\label{sec:SI_formalism}

We summarize here a formalism detailed fully in \cite{photoeccentric}, and describe how we employ it in our lightcurve modeling procedure. With a fortuitous rearrangement of Newton's version of Kepler's third law \cite{seager_unique_2003}, we can express a directly measured quantity from the transit lightcurve (the ratio of the semimajor axis $a$ to the stellar radius $R_{\star}$) in terms of the planetary period $P$ and the mass and radius of the star $M_{\star}$ and $R_{\star}$. For illustrative purposes, we manipulate $M_{\star}$ and $R_{\star}$ to substitute $\rho_{\star}$ and find

\begin{equation}
    a/R_{\star} = \sqrt[3]{\frac{G M_{\star} P^{2}}{4\pi^2{R_{\star}}^3}}=\sqrt[3]{\frac{G P^{2}}{3\pi{\rho_{\star}}}}.
    \label{eq:SI_kepler}
\end{equation}

In the simplified case of a circular orbit with an impact parameter $b=0$, the transit duration is given by the time to sweep across $2R_{\star}$. The planet's speed is constant in this case, $2\pi a/P$, so that the duration can be very roughly approximated at $T\sim P/\pi (a/R_{\star}) $. However, since the effect of a non-zero impact parameter may influence the transit duration in a similar way as a non-zero eccentricity, we must incorporate these variables into our expression for transit duration. Including the additional complexities of an eccentric orbit characterized by an eccentricity $e$, longitude of periapse $\omega$, inclination $i$ and transit depth $\delta$, the full expression for transit duration is given by Equation \ref{eq:SI_T14}:  

\begin{dmath}
    T_{14/23} = \frac{P}{\pi}\frac{(1-e^2)^{3/2}}{(1+e\sin\omega)^2} \arcsin \left( \frac{ \sqrt{ (1 +/- \delta^{1/2})^2 - (a/R_{\star})^2 \left( \frac{1-e^2}{1+e\sin w} \right)^2 \cos^2 i}}{(a/R_{\star})\frac{1-e^2}{1+e \sin w} \sin i} \right),
    \label{eq:SI_T14}
\end{dmath}

where $T_{14}$ is the duration between first and fourth contact and $T_{23}$ is between 2nd and 3rd contact (see \cite{Winn10_chapter} for additional description). $T_{14}$ is calculated with $(1 + \delta^{1/2})$, and $T_{23}$ is calculated with $(1 - \delta^{1/2})$. 

We include this analytic expression (taken from \cite{photoeccentric}) with the understanding that it reflects an approximation without accounting for limb darkening. We wish to convey the formalism of the relationship between transit duration and eccentricity. When the effects of limb darkening are significant, this approximation is still useful, though it would underpredict the covariances and uncertainties of the fit parameters \cite{carter_2008}. Therefore, in our analysis, we fit for the limb darkening along with the other transit parameters, including the eccentricity directly.

\cite{photoeccentric} derived an expression involving $T_{14}$, $T_{23}$ and equation \ref{eq:SI_kepler} as follows:

\begin{dmath}
    \rho_{\star} = g^{-3} \biggl ( \frac{2 \delta^{1/4}}{\sqrt{T_{14}^2 - T_{23}^2}} \biggr )^3  \frac{3 P}{G\pi^2},
    \label{eq:SI_rhostar}
\end{dmath}

where $g$ is defined to be

\begin{equation}
    g(e,\omega) = \frac{1 + e \sin\omega}{\sqrt{1-e^2}}
    \label{eq:SI_gdef}
\end{equation}

In this way, $\rho_{\star}$ is related to two quantities: a quantity dependent entirely on observables from the transit ($T_{14}$, $T_{23}$, $\delta$, and $P$), and a quantity $g$ that separately encodes eccentricity information. With prior information about  $\rho_{star}$ from some other means (asteroseismology or spectroscopy), $g$ is in principle extractable. 

We continue to adopt the symbolism and formalism of \cite{photoeccentric} to describe the Bayesian statistical framework of our analysis. We describe a model lightcurve parameterized by $e$, $\omega$, $\rho_{\star}$, and $X$. $X$ represents all other parameters of the model light curve (such as orbital period, transit epoch, radius ratio, limb-darkening parameters, and impact parameter). We take the variable $D$ to represent the light curve data. We intend to determine the probability of various values of e and omega given the data. According to Bayes' theorem,

\begin{equation}
    P(e, \omega, \rho_{\star}, X | D) \propto P(D | e, \omega, \rho_{\star}, X) P(e, \omega, \rho_{\star}, X)
\end{equation}
where the last term $P(e, \omega, \rho_{\star}, X)$ represents the prior knowledge. We impose a non-uniform prior only on $\rho_{\star}$ based on the stellar densities and uncertainties we calculated above, from stellar parameters measured independently. We rewrite the probability as

\begin{equation}
    P(e, \omega, \rho_{\star}, X | D) \propto P(D | e, \omega, \rho_{\star}, X) P(\rho_{\star})
\end{equation}
To obtain a two-dimensional joint posterior distribution for e and omega, we marginalize over X and $\rho_{\star}$ and obtain

\begin{equation}
    P(e, \omega | D) \propto \int \int P(D | e, \omega, \rho_{\star}, X) P(\rho_{\star}) dX d\rho_{\star}
\end{equation}
And finally, we marginalize over $\omega$ to obtain

\begin{equation}
    P(e | D) \propto \int \int \int P(D | e, \omega, \rho_{\star}, X) P(\rho_{\star}) dX d\rho_{\star} d\omega
\end{equation}
We demonstrate that, when incorporating a Bayesian sampling method to explore parameter space, we are able to translate a prior on stellar density and uniform priors on $e$ and $\omega$ into a constraint on a planet's eccentricity.

We perform the lightcurve modeling using gradient descent with the \texttt{exoplanet} \cite{exoplanet:joss} and \texttt{pymc3} \cite{exoplanet:pymc3} packages. The free parameters are the orbital period $P$, transit epoch $t_0$, planet-star radius ratio \rprs, impact parameter $b$, quadratic limb-darkening parameters $u_{1}$ and $u_{2}$ (sampled uniformly across $q_{1}$ and $q_{2}$ using the triangular limb darkening parameterization of \cite{kipping_2013}, eccentricity parameters \sesinw and \secosw, and the stellar density $\rho_{\star}$. With \texttt{exoplanet}, the stellar density itself may be taken as a free parameter, and the transit light curve is modeled based on the combination of each sampled $\rho_{\star}$ and $P$. Therefore, it is not necessary to e.g. manually convert a prior on $\rho_{\star}$ to a prior on $a/R_{\star}$. Using each combination of $e$, $\omega$, and $b$, we calculate $\rho_{\star}$. We apply a normal prior on the free parameter $\rho$ centered at the calculated \rhos with $\sigma = \sigma_{\rho_{\star}}$, where $\sigma_{\rho_{\star}}$ is the calculated uncertainty for each star.

The priors for each parameter are listed in Table \ref{tab:SI_priors}. We first calculated the maximum a posteriori (MAP) model solution, and used the MAP solution to initialize the sampler. We sampled the model parameters using No-U Turn Sampling (NUTS) \cite{10.5555/2627435.2638586} with two chains 20,000 tuning steps and 20,000 posterior draws each. The sample acceptance rate is greater than $90\%$ for all fits. We calculate the Gelman-Rubin $\hat{R}$ statistic for each transit fit to check for convergence, and we find that $\hat{R} < 1.05$ for each parameter in each transit fit. The full posteriors, convergence statistics, corner plots and trace plots are available in the Data Supplement.

\begin{table}
\centering
\begin{tabular}{ccc}
\hline
 Free Parameter & Prior Distribution & Values \\
 \hline
 $P$ & Uniform & [$Period-0.1$, $Period+0.1$] (days) \\ 
 $t_0$ & Uniform & $[Epoch-0.1, Epoch+0.1]$ (days) \\ 
 \rprs & Uniform & $[0.0, 0.2]$ \\
 $b$ & Uniform & $[-1.2, 1.2]$ \\
 $u_{1}$ & Normal & $[u_1, 0.05]$ \\
 $u_{2}$ & Normal & $[u_2, 0.05]$ \\
 \sesinw & Uniform & $(-1, 1)$ \\
 \secosw & Uniform & $(-1, 1)$ \\
 $\rho$ & Normal & $[\rho_{\star}, \sigma_{\rho_{\star}}]$ \\
 \vspace{0.01cm}
\end{tabular}
\caption{Transit fit free parameter prior distributions and values. $Period$ and $Epoch$ in the Values column represent the published values for the associated parameters in the NASA Exoplanet Archive \citep{KOICumulative}. $u_1$ and $u_2$ represent the limb darkening coefficients published by \cite{claret_2011} for the respective host star. The limb darkening coefficients are sampled using the triangular limb darkening parameterization of \cite{kipping_2013}.}
\label{tab:SI_priors}
\end{table}

\subsection*{Application of Pipeline to \textit{Kepler} Sample} 
\label{sec:SI_real}

With the proof-of-concept described in the Appendix in hand, we apply the photoeccentric pipeline to the sample of \ntargs \textit{Kepler} planets. For multi-planet systems, we fit each planet individually. We discard any simultaneous transits. We create stitched light curves containing only the transits of a single planet in each system following the method in the Injection and Recovery Demonstration section of the SI Appendix. While fitting planets in multi-planet systems individually does not force common stellar density and limb darkening posteriors, this method simplifies the process of removing planets or false positives in multi-planet systems from our sample and significantly reduces the required computational resources. As a test, we performed a joint planet fit on the three-planet system Kepler-445 (KOI 2704) and compared the posteriors for the individual and system fits. We found the differences in the posteriors, including those for stellar density and limb darkening, to be marginal. All fit parameters were consistent well within $1\sigma$ between the joint and individual fits. \cite{mann_gold_2017} performed a similar test with the Kepler-42 (KOI 961) system and also found the differences in posteriors to be marginal.

If a system exhibits TTVs, fitting a transit light curve without correcting for TTVs may cause a "smeared" model fit that misrepresents the impact parameter \cite{kippingasteroseismology}. An inflated impact parameter may be compensated by an inflated transit duration, which may incorrectly suggest that the planet is eccentric \cite{vaneylen2015}. For systems with TTVs, we fit the Kepler light curve with \texttt{exoplanet} simultaneously with each individual transit time. We use the transit times published by \cite{thompson_2018} to set normal priors around each transit time $t_n$, with $\sigma_{tn} = 0.05 $ days (1.2 hours). Consequently, the free parameters for these systems do not include $P$ or $t_{0}$. According to \cite{swift_characterizing_2015}, KOIs 248.01, 248.02, 250.01, 250.02, 314.01, 314.02, 314.03, 886.01, 886.02, and 898.01, and 952.02 show evidence for transit timing variations (TTVs). Additionally, \cite{mazeh_2013} reported evidence for TTVs in KOI 902.01, which was not included in the sample of \cite{swift_characterizing_2015}. We fit transit times for KOI 902.01 as well.  \cite{thompson_2018} did not publish transit times and TTVs for KOI 898.01. In the interest of consistency, we do not fit TTVs for KOI 898.01. We do not see evidence of "smearing" or an inflated transit duration in the fit for KOI 898.01, so we conclude that a periodic transit model is appropriate for this planet and dataset.

We warn of a risk that low-amplitude TTVs may be undetected, or that the known TTVs are not sufficiently accounted for, affecting the eccentricity posteriors reported in this work. We compare the eccentricity distribution of planets with known TTVs and without TTVs, and we find no considerable differences between the two distributions. We also calculate $g$ (Equation \ref{eq:SI_gdef}) for each planet, and we find that values of $g$ greater than 1 and less than 1 are roughly equally common in our sample, suggesting that TTVs are sufficiently accouted for. The sub-sample of planets with known TTVs is small (11 planets), and all conclusions in this paper would be upheld if planets with known TTVs were removed from our sample.

For KOIs 255.02, 676.01, 676.02, 898.03, 936.01, 952.04, 961.03, 1427.01, 1427.02, 2704.03, 2715.02, 2715.03, 2842.02, 2842.03, 2926.03, and 2926.05, we do not fit the orbital period or epoch to reduce the computational needs of their fits. We instead fix the period and epoch to the value published in \cite{KOICumulative}. Because the orbital period posteriors for our sample tend to have uncertainties less than $10^{-4}$ days, we do not expect fixing the period to significantly affect the resulting transit fit posteriors.

As a test, we compared transit fit posteriors using long cadence and short cadence data for several planets in our sample. We find that though the long-cadence fits constrain parameters more loosely than the short-cadence fits, the eccentricity posteriors are consistent with one another. We contend that for KOIs where only long-cadence data are available, the eccentricity posteriors may be poorly constrained, but are not significantly biased.

\section*{Likelihood Functions for Underlying Eccentricity Models}

The complete likelihood functions we used for the underlying eccentricity models are as follows:

\begin{equation}
    p(obs|\theta) = \frac{1}{N} \prod_{k=1}^{K} \sum_{n=1}^{N} \frac{e_{k}^{n}}{\sigma^2} \exp \left( \frac{-{e_{k}^{n}}^2}{2 \sigma^2} \right) \left( \frac{1-e^2}{1 + e \sin \omega} \right)
    \label{eq:SI_rayleighpdf}
\end{equation}
for the Rayleigh distribution with parameter $\sigma$;
\begin{equation}
    p(obs|\theta) = \frac{1}{N} \prod_{k=1}^{K} \sum_{n=1}^{N} \frac{\exp{ \left( \frac{-e_{k}^{n}}{2\sigma} \right) }^{2}} {\sigma \sqrt{2 \pi}} \left( \frac{1-e^2}{1 + e \sin \omega} \right)
    \label{eq:SI_hgausspdf}
\end{equation}
for the half-Gaussian distribution with parameter $\sigma$; and
\begin{equation}
    p(obs|\theta) = \frac{1}{N} \prod_{k=1}^{K} \sum_{n=1}^{N} \frac{\Gamma(a + b) {(e_{k}^{n}})^{a-1} {(1-{e_{k}^{n}})}^{b-1}}{\Gamma(a) \Gamma(b)} \left( \frac{1-e^2}{1 + e \sin \omega} \right)
    \label{eq:SI_betapdf}
\end{equation}
for the Beta distribution with parameters $a$ and $b$.

\section*{Discussion of Stellar Metallicity Provenances}

We take stellar metallicities from Kepler DR25 as compiled by \cite{mathur_revised_2017} to calculate densities for each star. \cite{mathur_revised_2017} compiled stellar parameters for Kepler stars from several different sources. For the vast majority of our sample (124 stars), we take metallicities derived spectroscopically from \cite{muirhead_2014}. The next largest fraction (19 stars) have metallicities derived spectroscopically from \cite{muirhead_2012a}. The remainder of our sample has spectroscopically derived metallicities from the following sources: 2 stars from \cite{mann_2013a, mann_2013b}; 1 star from \cite{torres_2017}; and 6 stars from \cite{muirhead_2015}. Finally, we include photometrically derived metallicities from the following sources: 8 stars from \cite{pinsonneault_2012}, 2 stars from \cite{huber_2014}; and 2 stars from \cite{brown_2011}.

For the two largest stellar provenances (\cite{muirhead_2014} and \cite{muirhead_2012a}), we calculate the normalized transit duration ($T_{14} / P^3$) for planets in each subsample. We compare the total transit duration distributions in these two subsamples using a CDF plot (Figure) to ensure that the medians of both distributions appear similar. Because the vast majority of our sample has metallicities taken from one source, and over $93\%$ of our sample has metallicities from spectroscopy, we contend that taking stellar parameters from different sources has not significantly biased our results, especially considering we ultimately take these values from the Kepler DR25 catalog, designed to be used uniformly to support the final Kepler transit detection run \cite{mathur_revised_2017}.

\begin{figure}[p!]
    \centering
    \includegraphics[width=\textwidth]{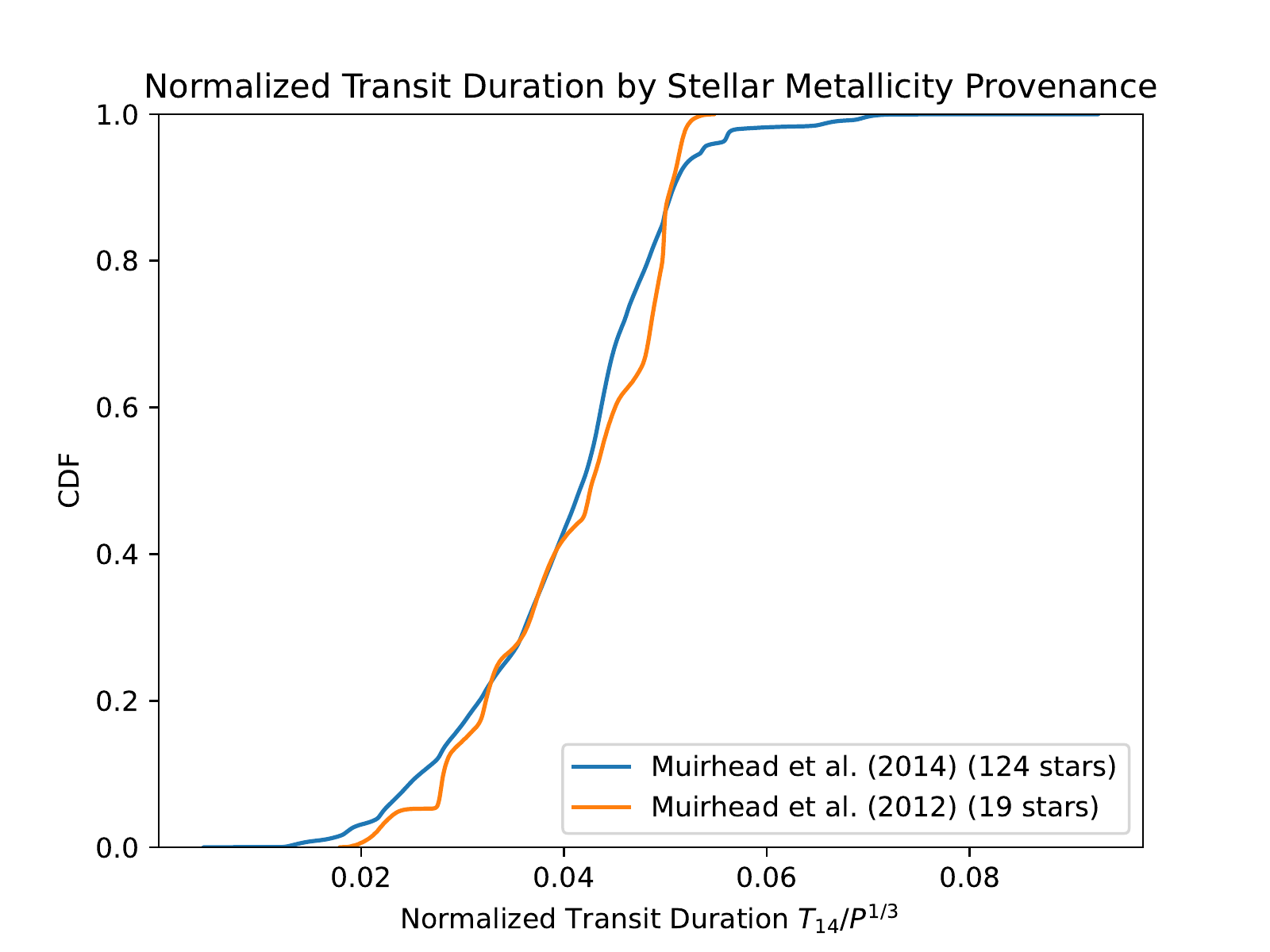}
    \caption{CDFs of normalized transit durations ($T_{14} / P^3$) for planets orbiting stars with metallicities taken from \cite{muirhead_2014} (blue) and for planets orbiting stars with metallicities taken from \cite{muirhead_2012a}. Stellar metallicities were taken from these sources via the Kepler DR25 catalog \cite{mathur_revised_2017}.}
    \label{fig:SI_metcdf}
\end{figure}

\section*{Injection and Recovery Demonstration}
\label{sec:SI_proof1}

To measure orbital eccentricities, we fit each planet's Kepler light curves using the calculated stellar density prior. We  perform an injection and recovery test simulating this procedure to ensure our pipeline accurately recovers known planetary properties, and to investigate any variability in detection sensitivity ($e$) and longitude of periastron ($\omega$) space. We simulate a suite of light curves of various signal-to-noise ratios (SNR). We define the SNR as 

\begin{equation}
    SNR = \frac{A}{\sigma} \sqrt{N*N_{t}}
    \label{eq:SI_SNR}
\end{equation}

where $A$ is the normalized transit depth, $\sigma$ is the individual flux error, $N$ is the number of observations in each transit, and $N_{t}$ is the number of transits in each light curve. We calculate $N$ by dividing the full transit duration by the appropriate flux cadence, and rounding to the nearest integer.

To investigate detection sensitivity across $e$ and $\omega$ space, we drew random combinations of $e$ and $\omega$ for several values of SNR. We drew between 200 and 500 combinations each for impact parameters of 0, 0.3, 0.6, and 0.8. $e$ and $\omega$ were drawn from uniform distributions with bounds $e = [0.0, 0.95]$, $\omega = [0.0, 360)$ degrees. We do not allow the injected or fit $e$ to be larger than 0.95. We draw a set of combinations for light curve SNRs of 10, 50, and 100, using Equation \ref{eq:SI_SNR} to calculate the corresponding flux error for individual points. We choose these values of SNR and $b$ to reflect the properties of our sample, the majority of which have transit SNRs between 10 and 70 according to \cite{KOICumulative}.

We calculate $N$ by first calculating the total transit duration ($T_{14}$) using Equation \ref{eq:SI_T14}, based on the injected values of e, omega, and impact parameter, and dividing $T_{14}$ by the appropriate flux cadence (one minute for short cadence and 30 minutes for long cadence). Therefore, the magnitude of flux error bars for each injected light curve is slightly different, based on the transit duration as determined by the injected $e$, $\omega$, and impact parameter, preserving the SNRs (and not necessarily the magnitude of the light curve uncertainties).

For each combination of $e$, $\omega$, $b$, and SNR, we create a light curve using these properties based on the transit properties of KOI 255.01. We chose KOI 255.01 to model the synthetic light curves because its transit properties are typical of our sample. The simulated light curves all have the same orbital period $P = 27.5$ days, planet/star radius ratio $R_p/R_{\star} = 0.044$, and quadratic limb darkening parameters $u_1 = 0.42$ and $u_2 = 0.30$. Each synthetic light curve necessarily has $N_{t} = 12$ transits per light curve. We obtain the planet parameters from the NASA Exoplanet Archive \cite{KOICumulative}. 

The simulated light curves all have the same semimajor axis/stellar radius ratio $a/R_{\star}$, which we calculate with Equation \ref{eq:SI_kepler}, where $M_{\star}$ and $R_{\star}$ are the stellar mass and radius calculated using the method in the Calculation of Stellar Densities section of the SI Appendix, respectively. We calculate $a/R_{\star}$ rather than using the values published by \cite{KOICumulative} to ensure the system is consistent with our calculated mass and radius. Using each combination of $e$, $\omega$, and $b$, we calculate $\rho_{\star}$ for each simulated system. We create simulated light curves using the transit modeling Python package \texttt{batman} \cite{batman}. This process is repeated twice for simulated short-cadence and long-cadence data. For long-cadence light curves, each flux point is integrated over an exposure time of 30 minutes. We process the simulated light curves according to the procedure described in the Lightcurve Preparation section of the SI Appendix. To fit the transits, we apply a normal prior on the free parameter $\rho$ centered at the calculated \rhos with $\sigma = 1 \rho_{\odot}$. A prior with $\sigma = 1 \rho_{\odot}$ reflects a typical width of stellar density priors in our sample. We sampled the model parameters using No-U Turn Sampling (NUTS) \cite{10.5555/2627435.2638586} with chains of at least 1,000 draws each. We use 1,000 tuning steps for each fit.

We analyze the accuracy of the recovered eccentricity and longitude of periastron parameters in ($\sqrt{e} sin \omega$, $\sqrt{e} cos \omega$) space. For each injected and recovered transit, we calculate the sensitivity metric $N_\sigma$, where

\begin{equation}
    N_\sigma = \left\langle \frac{\text{abs}({\sqrt{e} \sin \omega}_{inj} - {\sqrt{e} \sin \omega}_{rec})}{\sigma_{e \sin \omega}}, \frac{\text{abs}({\sqrt{e} \cos \omega}_{inj} - {\sqrt{e} \cos \omega}_{rec})}{\sigma_{e \cos \omega}} \right\rangle
\end{equation}

where $inj$ refers to the injected parameter, $rec$ refers to the recovered parameter, and $\sigma$ is the standard deviation of the ${\sqrt{e} \sin \omega}$ or ${\sqrt{e} \cos \omega}$ posterior. We take the point estimates ${\sqrt{e} \sin \omega}_{rec}$ and ${\sqrt{e} \cos \omega}_{rec}$ to be the mean of the respective posterior. This metric represents the number of posterior standard deviations a recovered value falls from the injected value.

\begin{figure*}[p!]
  \includegraphics[width=1.0\linewidth]{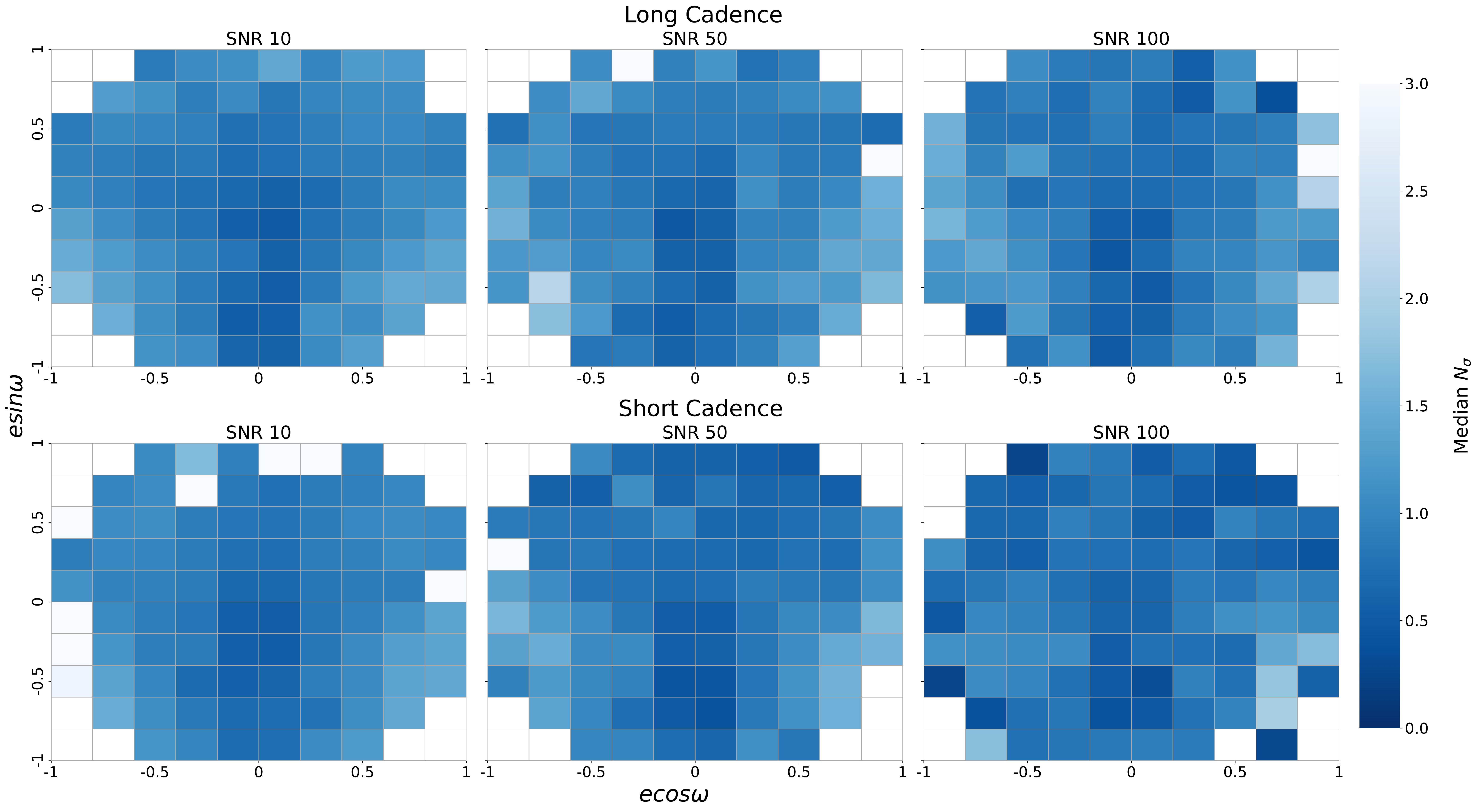}
  \caption{Injection and recovery sensitivity in ($e cos \omega$, $e sin \omega$) space across SNRs of 10, 50, and 100 for simulated Kepler long- and short-cadence light curves. The color bar represents the number of standard deviations the recovered ($e cos \omega$, $e sin \omega$) lies from the injected ($e cos \omega$, $e sin \omega$). The mean error is shown for the simulated transits that fall in each bin. Simulated transit parameters were drawn from uniform $e$ and $\omega$ distributions and sampled uniformly in $(\sqrt{e} cos \omega, \sqrt{e} sin \omega)$ space. The sensitivity is approximately uniform in parameter space with little dependence on SNR.}
  \label{fig:SI_snr}
\end{figure*}

Figure \ref{fig:SI_snr} shows the sensitivity in parameter space for long cadence and short cadence data in (${e \cos \omega}$, ${e \sin \omega}$) space for injected SNRs of 10, 50, and 100. The color bar represents the median value of $N_{\sigma}$ in each bin. The sensitivity is approximately uniform in parameter space, and all bins have a mean error less than $2.5 \sigma$. We demonstrate that the transit fitting machinery accurately recovers eccentricities using the light curve transit duration and stellar density for both long- and short-cadence simulated Kepler data, with little dependence on transit signal-to-noise.

Figure \ref{fig:SI_snr} shows the sensitivity in parameter space for long cadence and short cadence data, for injected SNRs of 10, 50, and 100. The color bar represents the median value of $N_{\sigma}$ in each bin. The sensitivity is approximately uniform in parameter space, and all bins have a mean error less than $2.5 \sigma$. We demonstrate that the transit fitting machinery accurately recovers eccentricities using the light curve transit duration and stellar density for both long- and short-cadence simulated Kepler data, with little dependence on transit signal-to-noise.

\subsection*{Inference of Simulated Parent Distribution} 
\label{subsec:SI_SimHBM}

We employ an injection and recovery technique to verify our hierarchical Bayesian inference technique to draw out the underlying eccentricity distribution. We designate \textit{a priori} a functional form for this distribution, and then draw from it to assign eccentricities to a synthetic planetary sample. We ought ideally then to recover this distribution, if we properly account for selection bias. 

 We performed six injection and recovery simulations, drawing from three functional forms for eccentricity. We employ the same set of functions as those tested by \cite{vaneylen} for the exoplanet eccentricity distribution. These include Rayleigh distributions with $\sigma = 0.2$ and $0.5$, Beta distributions with $a = 0.8, b = 3.0$ and $a = 2, b = 10$, and half-Gaussian distributions with $\sigma = 0.2$ and $0.5$. For each simulation, we repeat the steps outlined in the Injection and Recovery section of the SI Appendix using only short cadence data with an SNR of 100. We again model all simulated light curves based on the properties of KOI 255.10, with $P = 27.5$ days, planet/star radius ratio $R_p/R_{\star} = 0.044$, and quadratic limb darkening parameters $u_1 = 0.42$ and $u_2 = 0.30$.

Instead of assigning one of three impact parameters to each simulated planet, as in the Injection and Recovery section of the SI Appendix, we randomly draw a point on a unit sphere for the orbital inclination. If the inclination produces an orbital path where a planet may not fully transit ($b > 0.9$), we discard the simulation. We reject with $b > 0.9$ because allowing the sampler to explore $b$ approaching $1$ for grazing transits greatly increases the computational resources needed to perform this simulation. We discarded draws with $b>0.9$ and only allowed the sampler to explore $b <= 0.9$ to mitigate this effect. Likewise, if we draw a combination of $e$ and $\omega$ that is not physical (e.g. the planet at periapse is less than 1 $R_{\star}$ away from the star), we discard the simulation. The discarded draws were not replaced with another draw, but we continue drawing until we reach 50 acceptable draws for each underlying distribution. We draw and simulate light curves for 50 $e$s in each underlying distribution.

\begin{figure*}
  \includegraphics[width=1.0\linewidth]{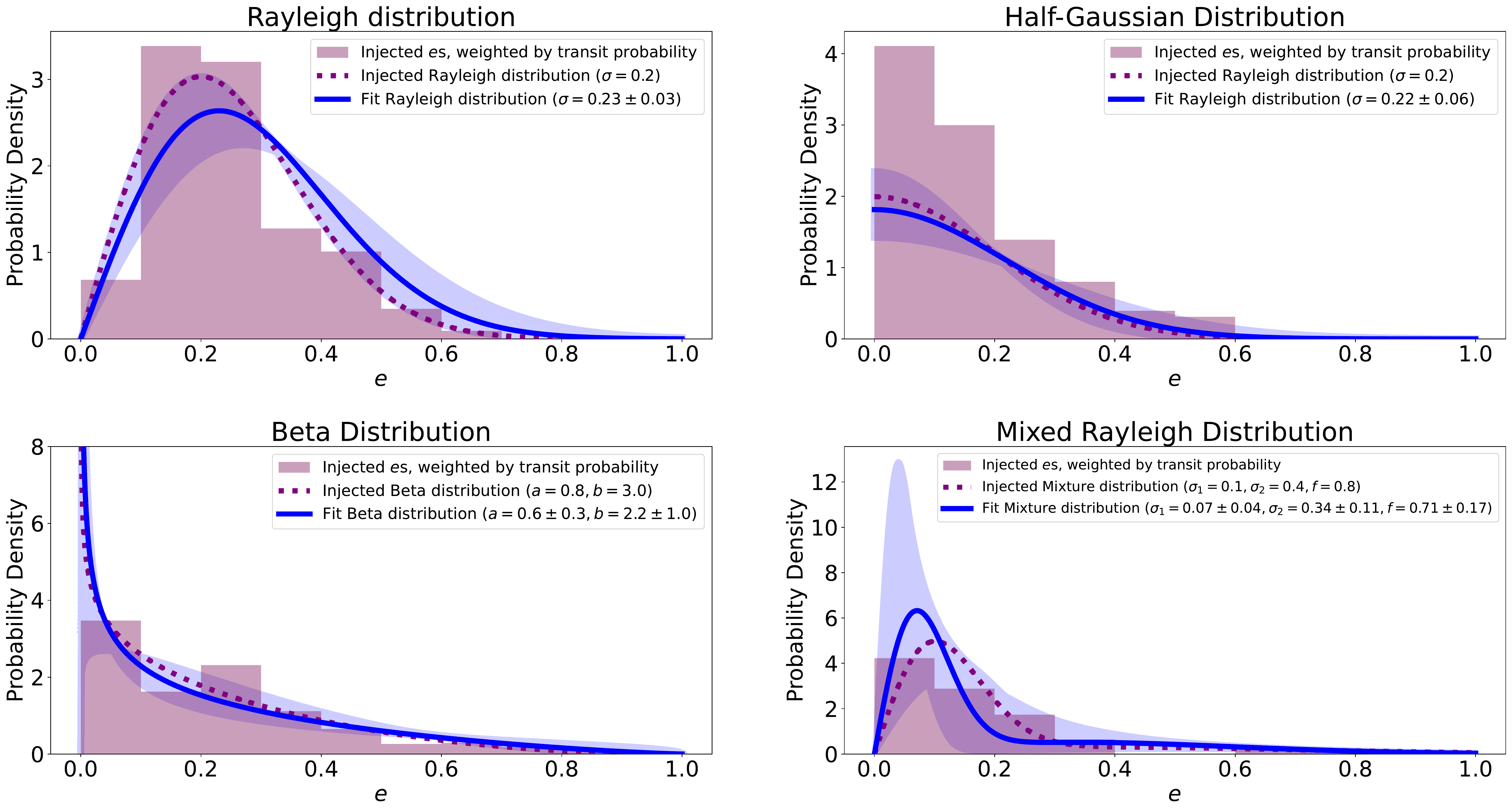}
  \caption{Injected and recovered Rayleigh, Half-Gaussian, Beta, and Mixture $e$ distributions. Each subplot shows the underlying distribution from which the $e$s were drawn (dotted purple), the histogram of drawn and injected $e$s (pink), and the recovered distribution (solid blue). The blue shaded regions denote the $1\sigma$ errors around the recovered distributions. }
  \label{fig:SI_simulatedHBI}
\end{figure*}

 We aim to recover the parameters of the single Rayleigh, Beta, and half-Gaussian distributions that we used to prescribe $e$ values for our synthetic sample. We randomly select 1000 points from the mock-up $e$ posterior distribution for each planet. We employ a uniform prior for all distribution parameters. We use a Markov Chain Monte Carlo (MCMC) analysis with the Python package \texttt{emcee} \cite{emcee}. The chains were run with 32 walkers for 2000 steps each, and we discarded a burn-in phase of 500 steps. We recover the true underlying $e$ distribution within 1-sigma in all cases. We show one injected and recovered distribution for each distribution type in Figure \ref{fig:SI_simulatedHBI}. In all cases, we recover the injected distribution parameters within $1\sigma$. 

We repeat this analysis for mixture models, specifically for a mixture model of Rayleigh distributions. We inject and recover two mixture model parameter sets: $\sigma_1 = 0.1, \sigma_2 = 0.4, f=0.8$ and $\sigma_1 = 0.25, \sigma_2 = 0.05, f=0.5$. In both cases, we recover the injected parameters within $1\sigma$. In Figure \ref{fig:SI_simulatedHBI}, we show the injected and recovered parameters for one of these simulations.

\section*{Transit Fit Corner Plots} 
\label{subsec:SI_corner}
\begin{figure*}
\centering
\begin{tabular}{cc}
  \includegraphics[width=90mm]{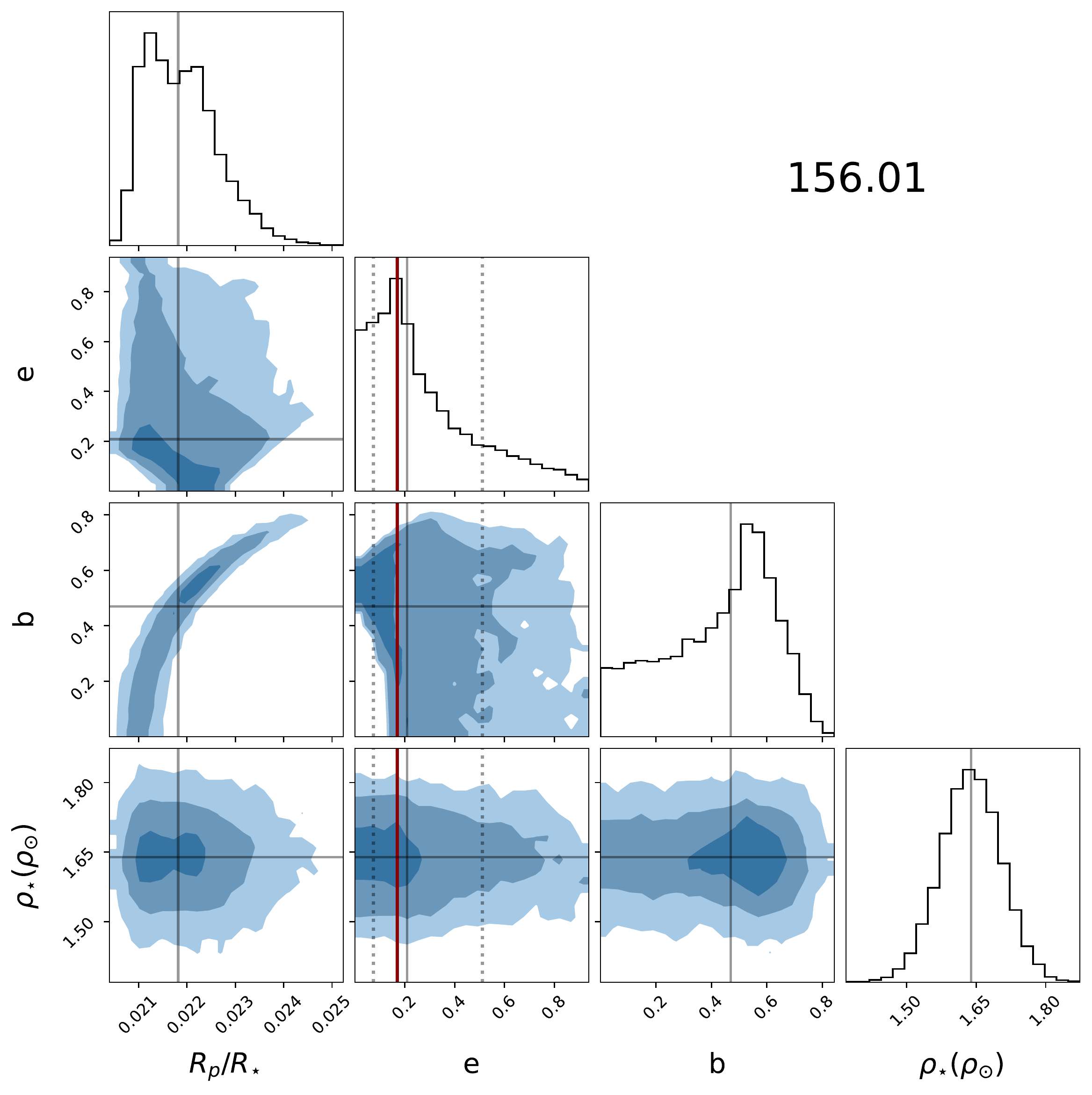} & \includegraphics[width=90mm]{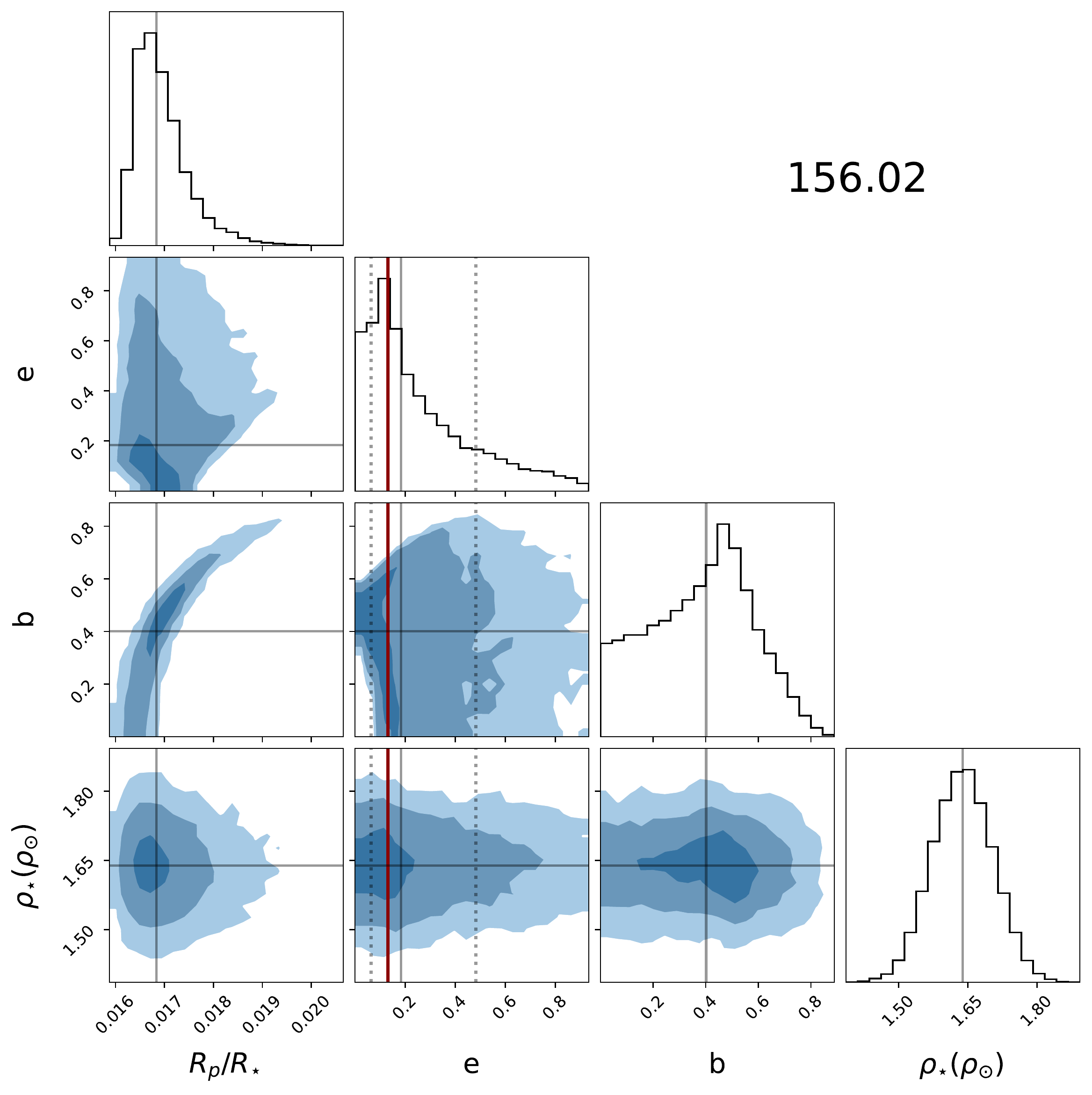} \\
 \includegraphics[width=90mm]{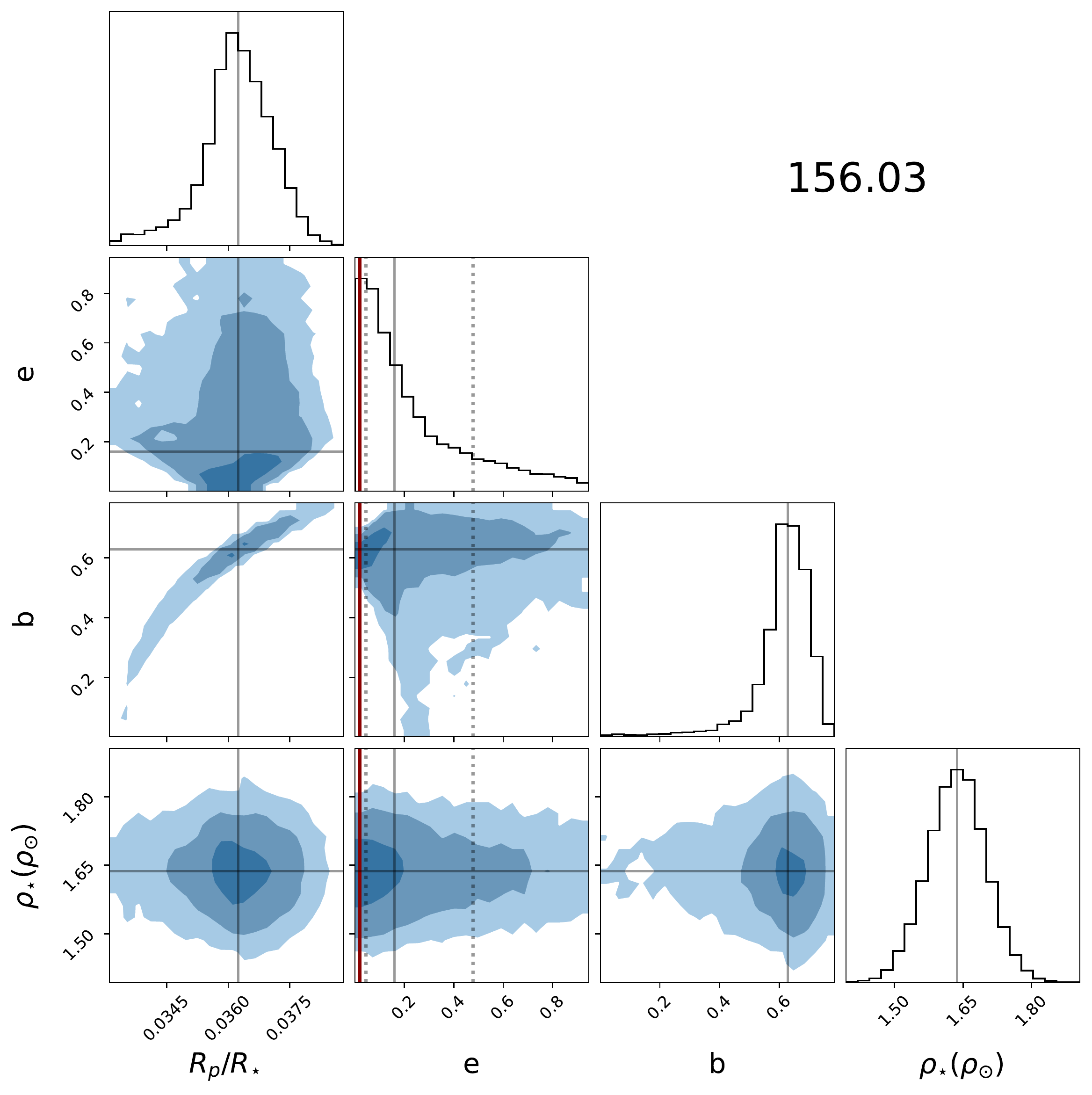} &   \includegraphics[width=90mm]{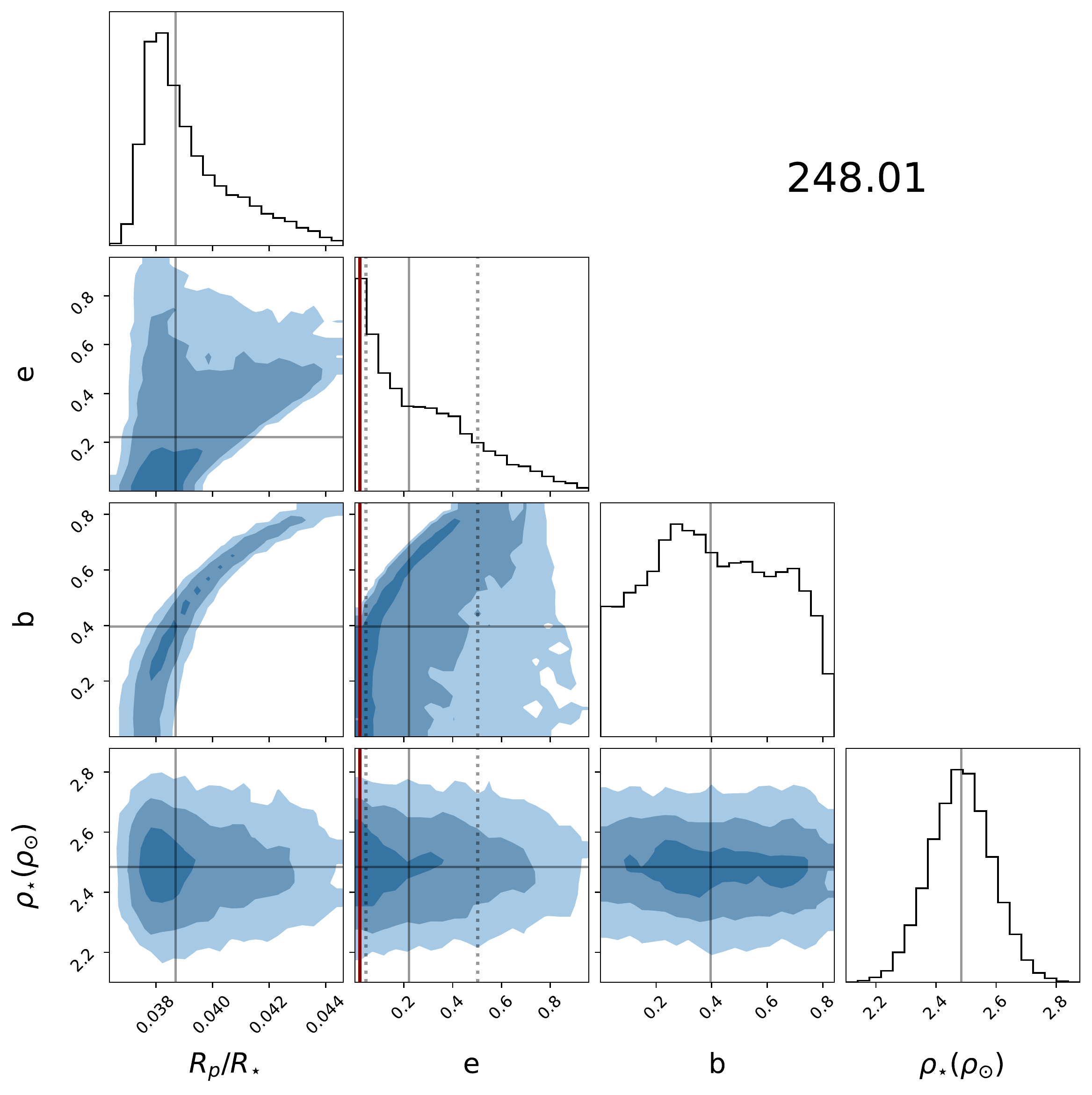} \\
\end{tabular}
\caption{We include the first four transit fit corner plots. Corner plots for the entire sample of M dwarf planets are available in the Data Supplements. The solid red line marks the statistical mode of the $e$ posterior. The solid gray line marks the median of the posteriors. The dotted gray lines mark the $16^{th}$ and $84^{th}$ percentiles of the $e$ posterior.}
\label{fig:SI_tfitcorner}
\end{figure*}

This section contains Figure \ref{fig:SI_tfitcorner}.
 
\bibliography{main}

\end{document}